\documentclass[sigplan,screen]{acmart}
\setcopyright{rightsretained}
\acmPrice{}
\acmDOI{10.1145/3519939.3523735}
\acmYear{2022}
\copyrightyear{2022}
\acmSubmissionID{pldi22main-p987-p}
\acmISBN{978-1-4503-9265-5/22/06}
\acmConference[PLDI '22]{Proceedings of the 43rd ACM SIGPLAN International Conference on Programming Language Design and Implementation}{June 13--17, 2022}{San Diego, CA, USA}
\acmBooktitle{Proceedings of the 43rd ACM SIGPLAN International Conference on Programming Language Design and Implementation (PLDI '22), June 13--17, 2022, San Diego, CA, USA}

\usepackage{balance}
\usepackage{booktabs}   
\usepackage{subcaption} 
\usepackage{algorithm}
\usepackage{algpseudocode}

\usepackage{mathpartir}
\usepackage{listings}
\usepackage{mathpartir}
\usepackage{amsmath,amsthm}
\usepackage{xspace}
\usepackage{tabularx}

\newcommand{\name}{\textsc{Peepul}\xspace}

\newenvironment{nop}{}{}

\definecolor{Bittersweet}{rgb}{1.0, 0.44, 0.37}
\definecolor{MidnightBlue}{rgb}{0.0, 0.2, 0.4}
\definecolor{BrightBlue}{rgb}{0.0, 0.2, 0.7}
\definecolor{byzantine}{rgb}{0.74, 0.2, 0.64}
\definecolor{caribbeangreen}{rgb}{0.0, 0.8, 0.6}

\lstMakeShortInline[columns=fullflexible]|

\begin{document}

\title{Certified Mergeable Replicated Data Types \\ (Extended Version)}


\author{Vimala Soundarapandian}
\affiliation{
	\institution{IIT Madras}
	\city{Chennai}
  \country{India}
}
\email{cs19d013@cse.iitm.ac.in}

\author{Adharsh Kamath}
\affiliation{
  \institution{NITK Surathkal}
	\city{Surathkal}
  \country{India}
}
\email{adharshkamathr@gmail.com}

\author{Kartik Nagar}
\affiliation{
	\institution{IIT Madras}
	\city{Chennai}
  \country{India}
}
\email{nagark@cse.iitm.ac.in}

\author{KC Sivaramakrishnan}
\affiliation{
	\institution{IIT Madras and Tarides}
	\city{Chennai}
  \country{India}
}
\email{kcsrk@cse.iitm.ac.in}

\begin{abstract}
	Replicated data types (RDTs) are data structures that permit concurrent
	modification of multiple, potentially geo-distributed, replicas without
	coordination between them. RDTs are designed in such a way that conflicting
	operations are eventually \emph{deterministically} reconciled ensuring
	\emph{convergence}. Constructing correct RDTs remains a difficult endeavour
	due to the complexity of reasoning about independently evolving states of the
	replicas. With the focus on the correctness of RDTs (and rightly so),
	existing approaches to RDTs are less efficient compared to their sequential
	counterparts in terms of time- and space-complexity of local operations. This
	is unfortunate since RDTs are often used in an local-first setting where the
	local operations far outweigh remote communication.

	In this paper, we present \name, a pragmatic approach to building and
	verifying efficient RDTs. To make reasoning about correctness easier, we cast
	RDTs in the mould of distributed version control system, and equip it with a
	three-way merge function for reconciling conflicting versions. Further, we go
	beyond just verifying convergence, and provide a methodology to verify
	arbitrarily complex specifications.  We develop a replication-aware
	simulation relation to relate RDT specifications to their efficient purely
	functional implementations. We have developed \name as an F* library that
	discharges proof obligations to an SMT solver. The verified efficient RDTs
	are extracted as OCaml code and used in Irmin, a Git-like distributed
	database.
\end{abstract}

\ccsdesc[500]{Computing methodologies~Distributed programming languages}
\ccsdesc[500]{Software and its engineering~Formal software verification}
\ccsdesc[500]{Computer systems organization~Availability}

\keywords{MRDTs, Eventual consistency, Automated verification, Replication-aware simulation}  

\maketitle

\section{Introduction}
\label{sec:intro}

Modern cloud-based software services often replicate data across multiple
geographically distributed locations in order to tolerate against partial
failures of servers and to minimise latency by bringing data closer to the
user. While services like Google Docs allow several users to concurrently edit
the document, the conflicts are resolved with the help of a centralised server.
On the other hand, services like Github and Gitlab, built on the decentralised
version control system Git, avoid the need for a centralised server, and permit
the different replicas (forks) to synchronize with each other in a peer-to-peer
fashion. By avoiding centralised server, \emph{local-first
software}~\cite{Kleppmann19} such as Git bring in additional benefits of
security, privacy and user ownership of data.

While Git is designed for line-based editing of text files and requires manual
intervention in the presence of merge conflicts, RDTs generalise this concept
to arbitrary general purpose data structures such as lists and hash maps, and
ensure convergence without manual intervention. Convergent Replicated Data
Types (CRDTs)~\cite{Shapiro}, which arose from distributed systems research,
are complete reimplementations of sequential counterparts aimed at providing
convergence without user intervention, and have been deployed in distributed
data bases such as AntidoteDB~\cite{Antidote} and Riak~\cite{Riak}.

In order to resolve conflicting updates, CRDTs generally need to carry their
causal contexts as metadata~\cite{Yu20}. Managing this causal context is often
expensive and complicated. For example, consider the observed-removed set CRDT
(OR-set)~\cite{Shapiro}, where, in the case of concurrent addition and removal,
the addition wins. A typical OR-set implementation uses two grow-only sets, one
for elements added to the set $\mathcal{A}$ and another for elements that are
removed $\mathcal{R}$. An element $e$ is removed from the OR-set by adding it
to the set $\mathcal{R}$, and thus creating a \emph{tombstone} for $e$. The set
membership is given by the difference between the two: $\mathcal{A} -
\mathcal{R}$, and two concurrent versions can be merged by unioning the
individual $\mathcal{A}$ and $\mathcal{R}$ sets. Observe that the tombstones
for removed elements cannot be garbage collected as that would require all the
replicas to remove the element at the same time, which requires global
coordination. This leads to an inefficient implementation. Several techniques
have been proposed to minimise this metadata overhead~\cite{Yu20,Delta}, but
the fundamental problem still remains.

\subsection{Mergeable Replicated Data Types}
\label{sec:mrdt}

As an alternative to CRDTs, mergeable replicated data types (MRDTs)~\cite{Kaki}
have been proposed, which extend the idea of distributed version control for
arbitrary data types. The causal context necessary for resolving the conflicts
is maintained by the MRDT middleware. MRDTs allow ordinary purely functional
data structures~\cite{okasaki_1998} to be promoted to RDTs by equipping them
with a three-way merge function that describes the conflict resolution policy.
When conflicting updates need to be reconciled, the causal history is used to
determine the lowest common ancestor (lca) for use in the three-way merge
function along with the conflicting states. The MRDT middleware garbage
collects the causal histories when appropriate~\cite{Dubey21}, and is no longer
a concern for the RDT library developer. This branch-consistent view of
replication not only makes it easier to develop individual data types, but also
leads to a natural transactional semantics~\cite{Tardis,Dubey20}.

An efficient OR-set MRDT that avoids tombstones can be implemented as follows.
We represent the OR-set as a list of pairs of the element and a unique id,
which is generated per operation. The list may have duplicate elements with
different ids. Adding an element appends the element and the id pair to the
head of the list ($O(1)$ operation). Removing an element removes all the
occurrences of the element from the list ($O(n)$ operation). Given two
concurrent versions of the OR-set $a$ and $b$, and their lowest common ancestor
$l$, the merge is implemented as $(a-l) ~@~ (b-l) ~@~ (l \cap a \cap b)$, where
$@$ stands for list append. Intuitively, we append the lists formed by newly
added elements in $a$ and $b$ with the list of elements that are present on all
the three versions. The unique id associated with the element ensures that in
the presence of concurrent addition and removal of the same element, the newly
added element with the fresh id, which has not been seen by the concurrent
remove, will remain in the merged result. The merge operation can be
implemented in $O(n~log~n)$ time by sorting the individual lists. In
\S\ref{sec:or_set_space}, we show how to make this implementation even more
efficient by removing the duplicate elements with different ids from the
OR-set.

\subsection{Efficiency and correctness}

The key question is how do we guarantee that such efficient implementations still preserve the intent of the OR-Set in a sound manner?
 Optimisations such as
removing duplicate elements are notoriously difficult to get right since the
replica states evolve independently. Moreover, individually correct RDTs may
fail to preserve convergence when put together~\cite{Kleppmann20}. Kaki et
al.~\cite{Kaki} opine that merge functions should not be written by hand, but
automatically derived from a relational representation of the sequential data
type. Their idea is to capture the key properties of the algebraic data type as
relations over its constituent elements. Then, the merge function devolves to a
merge of these relations (sets) expressed as MRDTs. During merge, the concrete
implementations are reified to their relational representations expressed in
terms of sets, merged using set semantics, and the final concrete state is
reconstructed from the relational set representation.

Unfortunately, mapping complex data types to sets does not lead to efficient
implementations. For example, a queue in Kaki et al. is represented by two
characteristic relations -- a unary relation for membership and a binary
relation for ordering. For a queue with $n$ elements, the ordering relation
contains $n^2$ elements. Reifying the queue to its characteristic relations and
back to its concrete representation for every merge is inefficient and
impractical. This technique does not scale as the structure of the data type
gets richer (Red-Black tree, JSON, file systems, etc.). The more complex the
data type, more complex the characteristic relations become, having an impact
on the cost of merge. Further Kaki et al. do not consider functional
correctness of MRDT implementations, but instead only focuses on the
correctness of convergence.

\subsection{Certified MRDTs}

Precisely specifying and verifying the functional correctness of efficient RDT
implementations is not straightforward due to the complexity of handling conflicts between divergent versions.
 This results in a huge
gap between the high-level specifications and efficient implementations. In
this work, we propose to bridge this gap by using Burckhardt et al.'s
\emph{replication-aware simulation relation}~\cite{Burckhardt}. However,
Burckhardt et al.'s simulation is only applicable to CRDTs and cannot be
directly extended to MRDTs which assume a different system model.

We first propose a system model and an operational semantics for MRDTs, and
precisely define the problem of convergence and functional correctness for
MRDTs. We also introduce a new notion of \emph{convergence modulo observable
behaviour}, which allows replicas to converge to different states, as long as
their observable behaviour to clients remains the same. This notion allows us
to build and verify even more efficient MRDTs.

Further, we go beyond Burckhardt et al.'s work \cite{Burckhardt} in the use of simulation relations by
mechanizing and automating (to an extent) the complete verification process. We
instantiate our technique as an F* library named \name and mechanically verify
the implementation of a number efficient purely functional MRDT implementations
including an efficient replicated two-list queue. Our replicated queue supports
constant time push and pop operations, a linear time merge operation, and does
not have any tombstones. To the best of our knowledge, ours is the first formal
declarative specification of a distributed queue (\S\ref{sec:queue}), and its
mechanised proof of correctness.

Being a SMT-solver-aided programming language, F* allows us to discharge many
of the proof obligations automatically through the SMT solver. Even though our
approach requires the simulation relation as input, we also observe that in
most cases, the simulation relation directly follows from the declarative
specification.

Our technique also supports composition, and we demonstrate how parametric
polymorphism allows composition of not just the MRDT implementations but also
their proofs of correctness. From our MRDT implementations in F*, we extract
verified OCaml implementations and execute them on top of Irmin, a Git-like
distributed database. Our experimental evaluation shows that our efficient MRDT
implementations scale significantly better than other RDT implementations.

To summarize, we make the following contributions:

\begin{itemize}
	\item We propose a store semantics for MRDT implementations and formally
		define the convergence and functional correctness problem for MRDTs,
		including a new notion of convergence modulo observable behaviour.
	\item We propose a technique to verify both convergence and functional
		correctness of MRDTs by adapting the notion of replication-aware simulation
		relation \cite{Burckhardt} to the MRDT setting.
	\item We mechanize and automate the complete verification process using F*,
		and apply our technique on a number of complex MRDT implementations,
		including a new time and space-efficient ORSet and a queue MRDT.
	\item We provide experimental results which demonstrate that our efficient
		MRDT implementations perform much better as compared with previous
		implementations, and also show the tradeoff between proof automation and
		verification time in F*.
\end{itemize}

The rest of the paper is organized as follows. \S\ref{sec:rdt} presents the
implementation model and the declarative specification framework for MRDTs.
\S\ref{sec:store} presents the formal semantics of the git-like replicated
store on which MRDT implementations run. In \S\ref{sec:correctness}, we present
a new verification strategy for MRDTs based on the notion of replication-aware
simulation. \S\ref{sec:comp} highlights the compositionality of our technique
in verifying complex verified MRDTs reusing the proofs of simpler ones.
\S\ref{sec:queue} presents the formally verified efficient replicated queue.
\S\ref{sec:results} presents the experimental evaluation and
\S\ref{sec:related} presents the related work.

\section{Implementing and Specifying MRDTs}
\label{sec:rdt}

In this section, we present the formal model for describing MRDT
implementations and their specifications.

\newcommand{\F}[1]{\mathsf{#1}}

\subsection{Implementation}
\label{sec:impl}

Our model of replicated datastore is similar to a distributed version control
system like Git~\cite{Git}, with replication centred around versioned states in
branches and explicit merges. A typical replicated datastore will have a
key-value interface with the capability to store arbitrary objects as
values~\cite{Riak,Irmin}. Since our goal is to verify correct implementations
of individual replicated objects, our formalism models a store with a single
object.

A replicated datastore consists of an object which is replicated across
multiple \textbf{\emph{branches}} $b_{1}, b_{2},\ldots \in branchID$. Clients
interact with the store by performing \textbf{\emph{operations}} on the object
at a specified branch, modifying its local state. The different branches may
concurrently update their local states and progress independently. We also
allow dynamic creation of a new branch by copying the state of an existing
branch. A branch at any time can get updates from any other branch by
performing a \textbf{\emph{merge}} with that branch, updating its local copy to
reflect the merge. Conflicts might arise when the same object is modified in
two or more branches, and these are resolved in an data type specific way.

An object has a type $\tau \in Type$, whose \textbf{\emph{type signature}}
$(Op_{\tau}, Val_{\tau})$ determines the set of supported operations
$Op_{\tau}$ and the set of their return values $Val_{\tau}$. A special value
$\bot \in Val_{\tau}$ is used for operations that return no value.

\begin{definition}
\label{def:type}
A \textbf{\emph{mergeable replicated data type (MRDT) implementation}} for a
	data type $\tau$ is a tuple $D_{\tau} = (\Sigma, \sigma_{0}, do,\allowbreak
	merge)$ where:
\begin{itemize}
	\item $\Sigma$ is the set of all possible states at a branch,
	\item $\sigma_{0} \in \Sigma$ is the initial state,
	\item $do : Op_{\tau} \times \Sigma \times Timestamp \rightarrow \Sigma
		\times Val_{\tau}$ implements every data type operation,
	\item $merge : \Sigma \times \Sigma \times \Sigma \rightarrow \Sigma$
		implements the three-way merge strategy.
\end{itemize}
\end{definition}

An MRDT implementation $\mathcal{D_{\tau}}$ provides two \emph{methods}:
\emph{do} and \emph{merge} that the datastore will invoke appropriately. We
assume that these methods execute atomically. A client request to perform an
operation $o \in Op_{\tau}$ at a branch triggers the call $do (o, \sigma, t)$.
This takes the current state $\sigma \in \Sigma$ of the object at the branch
where the request is issued and a timestamp $t \in Timestamp$ provided by the
datastore, and produces the updated object state and the return value of the
operation.

The datastore guarantees that the timestamps are unique across all of the
branches, and for any two operations $a$ and $b$, with timestamps $t_a$ and
$t_b$, if $a$ happens-before $b$, then $t_a < t_b$. The data type
implementation can use the timestamp provided to implement the
conflict-resolution strategy, but is also free to ignore it. For simplicity of
presentation, we assume that the timestamps are positive integers, $Timestamp = \mathbb{N}$.
The datastore may choose to implement the timestamp using Lamport
clocks~\cite{Lamport}, along with the unique branch id to provide uniqueness
of timestamps.

A branch $a$ may get updates from another branch $b$ by performing a merge,
which modifies the state of the object in branch $a$. In this case, the
datastore will invoke $merge(\sigma_{lca},\allowbreak \sigma_{a},\allowbreak
\sigma_{b})$ where $\sigma_{a}$ and $\sigma_{b}$ are the current states of
branch $a$ and $b$ respectively, and $\sigma_{lca}$ is the lowest common
ancestor (LCA) of the two branches. The LCA of two branches is the most recent
state from which the two branches diverged. We assume that execution of the
store will begin with a single branch, from which new branches may be
dynamically created. Hence, for any two branches, the LCA will always exist.

\subsubsection{OR-set}
\label{sec:or_set}

We illustrate MRDT implementations using the example of an OR-set. Recall from
\S\ref{sec:intro} that the OR-set favours the addition in the case where there
is a concurrent addition and removal of the same element on different branches.

\begin{figure}[ht]
\begin{algorithmic} [1]
\State $\Sigma = \mathcal{P}(\mathbb{N} \times \mathbb{N})$
\State $\sigma_{0} = \{\}$
\State $do(rd, \sigma, t) = (\sigma, \{a \mid (a,t) \in \sigma\})$
\State $do(add (a), \sigma, t) = (\sigma \cup \{(a,t)\} ,  \bot)$
\State $do(remove (a), \sigma, t) = (\{e \in \sigma \mid fst (e) \neq a\}, \bot)$
\State $merge( \sigma_{lca}, \sigma_a, \sigma_b) = $
	\Statex $(\sigma_{lca} \cap \sigma_a \cap \sigma_b) \cup (\sigma_a - \sigma_{lca}) \cup (\sigma_b - \sigma_{lca})$
\end{algorithmic}
\caption{OR-set data type implementation}
\label{fig:fig2}
\end{figure}

Let us assume that the elements in the OR-set are natural numbers. Its type
signature would be $(Op_{orset}, Val_{orset})$ $=(\{\F{add}(a),
\F{remove}(a)\mid a \in \mathbb{N}\} \cup \{\F{rd}\},
\{\mathcal{P}(\mathbb{N}),\bot\})$. Figure~\ref{fig:fig2} shows an MRDT
implementation of the OR-set data type. The state of the object is a set of
pairs of the element and the timestamp. The operations and the merge remain the
same as described in \S\ref{sec:mrdt}. Note that we use $fst$ and $snd$
functions to obtain the first and second elements respectively from a tuple.
This implementation may have duplicate entries of the same element with
different timestamps.

\subsubsection{Space-efficient OR-set (OR-set-space)}
\label{sec:or_set_space}

One possibility to make this OR-set implementation more space-efficient is by
removing the duplicate entries from the set. A duplicate element will appear in
the set if the client calls $\F{add}(e)$ for an element $e$ which is already in
the set. Can we reimplement $\F{add}$ such that we leave the set as is if the
set already has $e$? Unfortunately, this breaks the intent of the OR-set. In
particular, if there were a concurrent remove of $e$ on a different branch,
then $e$ will be removed when the branches are merged. The key insight is that
the effect of the duplicate add has to be recorded so as to not lose additions.

\begin{figure}[ht]
\begin{algorithmic} [1]
	\State $\Sigma = \mathcal{P}(\mathbb{N} \times \mathbb{N})$
	\State $\sigma_{0} = \{\}$
	\State $do(rd, \sigma, t) = (\sigma, \{a \mid (a,t) \in \sigma\})$
	\State $do(add(a), \sigma, t) = \text{if } (a, \_) \in \sigma \text{ then } (\sigma[a \mapsto t], \bot) $
		\State $\qquad \qquad \qquad \quad \text{ else } (\sigma \cup \{(a, t)\}, \bot)$
	\State $do(remove(a), \sigma, t) = (\{e \in \sigma \mid fst (e) \neq a\}, \bot)$

	\State $merge(\sigma_{lca}, \sigma_a, \sigma_b) =$
		\State $\quad \{e \mid e \in (\sigma_{lca} \cap \sigma_{a} \cap \sigma_{b})\} ~\cup$
		\State $\quad \{e \mid e \in (\sigma_{a} - \sigma_{lca}) \wedge (fst(e), \_) \notin (\sigma_{b} - \sigma_{lca}) \} ~\cup$
		\State $\quad \{e \mid e \in (\sigma_{b} - \sigma_{lca}) \wedge (fst(e), \_) \notin (\sigma_{a} - \sigma_{lca}) \} ~\cup$
		\State $\quad \{e \mid e \in (\sigma_{a} - \sigma_{lca}) ~\wedge$
		\State $\quad \quad \enskip \enskip (\forall t. ~(fst(e), t) \in (\sigma_{b} - \sigma_{lca}) \Rightarrow snd(e) > t) \} ~\cup$
		\State $\quad \{e \mid e \in (\sigma_{b} - \sigma_{lca}) ~\wedge$
		\State $\quad \quad \enskip \enskip (\forall t. ~(fst(e), t) \in (\sigma_{a} - \sigma_{lca}) \Rightarrow snd(e) > t) \}$
\end{algorithmic}
	\caption{Space-efficient OR-set (OR-set-space) implementation}
\label{fig:space_impl}
\end{figure}

Figure~\ref{fig:space_impl} provides the implementation of the space-efficient
OR-set. The read and the remove operations remain the same as the earlier
implementation. If the element being added is not already present in the set,
then the element is added to the set along with the timestamp. Otherwise, the
timestamp of the existing entry is updated to the new timestamp. Given that our
timestamps are unique, the new operation's timestamp will be distinct from the
old timestamp. This prevents a concurrent remove from deleting this new
addition.

Another possibility of duplicates is that the same element may concurrently be
added on two different branches. The implementation of the merge function now
has to take care of this possibility and not include duplicates. An element in
the merged set was either in the lca and the two concurrent states (line 8), or
was only added in one of the branches (lines 9 and 10), or was added in both
the branches in which case we pick the entry with the larger timestamp (lines
11--14).

\subsection{Specification}

Given that there are several candidates for implementing an MRDT, we need a way
to specify the behaviour of an MRDT so that we may ask the question of whether
the given implementation satisfies the specification. We now present a
declarative framework for specifying MRDTs which closely follows the framework
presented by Burckhardt et al.~\cite{Burckhardt}. We define our specifications
on an \emph{abstract state}, which capture the state of the distributed store.
It consists of \emph{events} in a execution of the distributed store, along
with a \emph{visibility} relation among them.

\begin{definition}
\label{def:abs}
An \textbf{\emph{abstract state}} for a data type $\tau=(Op_\tau,\allowbreak
	Val_\tau)$ is a tuple $I = \langle E, oper, rval, time, vis \rangle$, where
\begin{itemize}
	\item $E \subseteq Event$ is a set of events,
	\item $oper: E \rightarrow Op_{\tau}$ associates the data type operation with
		each event,
	\item $rval: E \rightarrow Val_{\tau}$ associates the return value with each
		event,
	\item $time: E \rightarrow Timestamp$ associates the timestamp at which an
		event was performed,
	\item $vis \subseteq E \times E$ is an irreflexive, asymmetric and transitive
		\textbf{\emph{visibility relation}}.
\end{itemize}
\end{definition}

Given $e \xrightarrow{vis} f$, $e$ is said to causally precede $f$. In our
setting, it may be the case that the operation of $f$ follows the operation of
$e$ on the same branch, or the operations of $f$ and $e$ were performed on
different branches $b_f$ and $b_e$, but before the operation of $f$, the branch
$b_e$ on which the operation of $e$ was performed was merged info $b_f$.

We specify a data type $\tau$ by a function $\mathcal{F_{\tau}}$ which
determines the return value of an operation $o$ based on prior operations
applied on that object. $\mathcal{F_{\tau}}$ also takes as a parameter the
abstract state that is visible to the operation. Note that the abstract state
contains all the information that is necessary to specify the return-value of
$o$.

\begin{definition}
A \textbf{\emph{replicated data type specification}} for a type $\tau$ is a
	function $\mathcal{F_{\tau}}$ that given an operation $o \in Op_\tau$ and an
	abstract state $I$ for $\tau$, specifies a return value $\mathcal{F_{\tau}}
	(o,I) \in Val_{\tau}$.
\end{definition}

\subsubsection{OR-set specification}
\label{sec:spec}

As an illustration of the specification language, let us consider the OR-set.
For the OR-set, both $\F{add}$ and $\F{remove}$ operations always return
$\bot$. We can formally specify the `add-wins' conflict resolution strategy as
follows:

\begin{equation*}
\label{eq:orset}
\begin{aligned}
	\mathcal{F}_{orset} (\F{rd}, \langle E, oper, rval, time, vis \rangle) = \{a \mid \exists e \in E.\ oper (e) \\= \F{add} (a) \wedge
			                       \neg (\exists f \in E.\ oper (f) = \F{remove} (a) \wedge e \xrightarrow{vis} f)\}
\end{aligned}
\end{equation*}

In words, the read operation returns all those elements for which there exists
an $\F{add}$ operation of the element which is not visible to a $\F{remove}$
operation of the same element. Hence, if an $\F{add}$ and $\F{remove}$
operation are concurrent, then the $\F{add}$ would win. Notice that the
specification, while precisely encoding the required semantics, is far removed
from the MRDT implementations of the OR-set that we saw earlier. Providing a
framework for bridging this gap in an automated and mechanized manner is one of
the principal contributions of this work.

\section{Store Semantics and MRDT Correctness}
\label{sec:store}

In this section, we formally define the semantics of a replicated datastore
$\mathbb{S}$ consisting of a single object with data type implementation
$\mathcal{D}_{\tau}$. Note that the store semantics can be easily generalized
to multiple objects (with possibly different data types), since the store
treats each object independently. We then define formally what it means for
data type implementations to satisfy their specifications. We also introduce a
novel notion of convergence across all the branches called \emph{convergence
modulo observable behaviour} that differs from the standard notions of eventual
consistency. This property allows us to have more efficient but verified
merges.

The \textbf{\emph{semantics}} of the store is a set of all its executions. In
order to easily relate the specifications which are in terms of abstract states
to the implementation, we maintain both the concrete state (as given by the
data type implementation) and the abstract state at every branch in our store
semantics. Formally, the semantics of the store are parametrised by a data type
$\tau$ and its implementation $D_\tau = (\Sigma, \sigma_0, do, merge)$. They
are represented by a labelled transition system $\mathcal{M}_{D_\tau} = (\Phi,
\rightarrow)$. Assume that $\mathcal{B}$ is the set of all possible branches.
Each state in $\Phi$ is a tuple $(\phi, \delta, t)$ where,

\begin{itemize}
	\item $\phi : \mathcal{B} \rightharpoonup \Sigma $ is a partial function that
		maps branches to their concrete states,
	\item $\delta : \mathcal{B} \rightharpoonup  I$ is a partial function that
		maps branches to their abstract states,
	\item $t \in Timestamp$ maintains the current timestamp to be supplied to
		operations.
\end{itemize}

The initial state of the labelled transition system consists of only one branch
$b_{\bot}$, and is represented by $C_{\bot}  = (\phi_{\bot}, \delta_{\bot}, 0)$
where $\phi_{\bot} = [b_{\bot} \mapsto \sigma_{0}]$ and $\delta_{\bot} =
[b_{\bot} \mapsto I_{0}]$.

Here, $\sigma_{0}$ is the initial state as given by the implementation
$D_\tau$, while $I_0$ is the empty abstract state, whose event set is empty. In
order to describe the transition rules, we first introduce abstract operations
$do^{\#}$, $merge^{\#}$ and $lca^{\#}$ which perform a data type operation,
merge operation and find the lowest common ancestor respectively on abstract
states:

\begin{algorithmic} [1]
	\Statex $do^{\#} \langle I,e,op,a,t \rangle$
	\Statex $\qquad = \langle I.E \cup \{e\}, I.oper [e \mapsto op], I.rval [e \mapsto a],$
	\Statex $\qquad \quad \; I.time [e \mapsto t], I.vis \cup \{(f,e) \mid f \in I.E)\} \rangle$
	\Statex
	\Statex $merge^{\#} (I_{a}, I_{b}) = I_m \text{ where}$
	\Statex $\qquad I_m.E = I_a.E \cup I_b.E$
	\Statex $\qquad \F{prop} \in \{oper, rval, time\}$
	\Statex $\qquad I_m.\F{prop}(e) =
		\begin{cases}
			I_a(e) \quad \text{if } e \in I_a.E\\
			I_b(e) \quad \text{if } e \in I_b.E
		\end{cases}$
	\Statex $\qquad I_m.vis = I_a.vis \cup I_b.vis$
	\Statex
	\Statex $lca^{\#}(I_a,I_b) = \langle I_a.E ~\cap~ I_b.E, I_a.oper\mid_{E_l}, $
	\Statex $\qquad \qquad \qquad I_a.rval\mid_{E_l}, I_a.time\mid_{E_l}, I_a.vis\mid_{E_l}\rangle$
	\Statex
\end{algorithmic}

In terms of abstract states, $do^{\#}$ simply adds the new event $e$ to the set
of events, appropriately setting the various event properties and visibility
relation. $merge^{\#}$ of two abstract states simply takes a union of the
events in the two states. Similarly, the $lca^{\#}$ of two abstract states
would be the intersection of events in the two states.

\begin{figure}[ht]
\centering
\[
\inferrule{b_{1} \in dom (\phi) \\ b_{2} \notin dom (\phi) \\  \phi^{'} = \phi [b_{2} \mapsto \phi(b_{1})] \\ \delta^{'} = \delta [b_{2} \mapsto \delta(b_{1})]}
          {(\phi, \delta, t) \xrightarrow{CREATEBRANCH (b_{1}, b_{2})} (\phi^{'}, \delta^{'}, t)\\}
\]

\[
\inferrule{b\in dom (\phi)  \\  \mathcal{D}_{\tau}.do (o, \phi (b), t) = (\sigma^{'}, a) \\ e : \{oper = o, time = t, rval = a\} \\
do^{\#} (\delta (b), e, o, a, t) = I^{'} \\ \phi^{'} = \phi [b \mapsto \sigma^{'}] \\ \delta^{'} = \delta [b \mapsto I^{'}]}
          {(\phi, \delta, t) \xrightarrow{DO (o, b)} (\phi^{'}, \delta^{'}, t+1)\\}
\]

\[
\inferrule{b_{1} \in dom (\phi) \\ b_{2} \in dom (\phi) \\ lca \in dom (\phi) \\ \delta(lca) = lca^{\#}(\delta(b_1), \delta(b_2)) \\
\mathcal{D}_{\tau}.merge (\phi(lca), \phi(b_{1}), \phi(b_{2})) = \sigma_{merge} \\
merge^{\#} (\delta (b_{1}), \delta (b_{2})) = I_{merge} \\
\phi^{'} = \phi [b_{1} \mapsto \sigma_{merge}] \\ \delta^{'} = \delta [b_{1} \mapsto I_{merge}]}
          {(\phi, \delta, t) \xrightarrow{MERGE (b_{1}, b_{2})} (\phi^{'}, \delta^{'}, t)}
\]
\caption{Semantics of the replicated datastore}
\label{fig:sem}
\end{figure}

Figure~\ref{fig:sem} describes the transition function $\rightarrow$. The first
rule describes the creation of new branch $b_{2}$ from the current branch
$b_{1}$. Both the concrete and abstract states of the new branch will be the
same as that of $b_{1}$. The second rule describes a branch $b$ performing an
operation $o$ which triggers a call to the $do$ method of the corresponding
data type implementation. The return value is recorded using the function
$rval$. A similar update is also performed on abstract state of branch $b$
using $do^{\#}$. The third rule describes the merging of branch $b_{2}$ into
branch $b_{1}$ which triggers a call to the $merge$ method of the data type
implementation. We assume that the store provides another branch $lca$ whose
abstract and concrete states correspond to the lowest common ancestor of the
two branches.

\begin{definition}
\label{def:exec}
An \textbf{\emph{execution}} $\chi$ of $\mathcal{M}_{D_\tau}$ is a finite but
	unbounded sequence of transitions starting from the initial state $C_{\bot}$.
\begin{equation}
\label{eq:2}
\chi = (\phi_{\bot}, \delta_{\bot}, 0) \xrightarrow{e_1} (\phi_{1}, \delta_{1}, t_{1}) \xrightarrow{e_2} \dots \xrightarrow{e_n} (\phi_{n}, \delta_{n}, t_{n})\\
\end{equation}
\end{definition}

\begin{definition}
\label{def:sat}
An execution $\chi$ \textbf{\emph{satisfies}} the specification
	$\mathcal{F}_{\tau}$ for the data type $\tau$, written as $\chi \models
	\mathcal{F}_{\tau}$, if for every $DO$ transition $(\phi_{i}, \delta_{i},
	t_{i}) \xrightarrow{DO (o, b)} (\phi_{i+1}, \delta_{i+1}, t_{i}+1)$ in
	$\chi$, such that $\mathcal{D}_{\tau}.do (o, \phi_{i} (b), t_{i}) = (\sigma,
	a)$, then $a = \mathcal{F}_{\tau} (o, \delta_i(b))$.
\end{definition}

That is for every operation $o$, the return value $a$ computed by the
implementation on the concrete state must be equal to the return value of the
specification function $\mathcal{F}_{\tau}$ computed on the abstract state.
Next we define the notion of convergence (i.e. strong eventual consistency) in
our setting:

\begin{definition}
\label{def:conv}
An execution $\chi$ (as in equation \ref{eq:2})  is \textbf{\emph{convergent}},
	\\if for every state $(\phi_{i}, \delta_{i})$ and
\begin{equation*}
\forall b_{1}, b_{2} \in dom (\phi_i). \delta_{i} (b_{1}) = \delta_{i} (b_{2}) \implies \phi_{i} (b_{1}) = \phi_{i} (b_{2})
\end{equation*}
\end{definition}

That is, two branches with the same abstract states--which corresponds to
having seen the same set of events--must also have the same concrete state. We
note that even though eventual consistency requires two branches to converge to
the same state, from the point of view of a client that uses the data store,
this state is never directly visible. That is, a client only notices the
operations and their return values. Based on this insight, we define the notion
of observational equivalence between states, and a new notion of convergence
modulo observable behaviour that requires branches to converge to states that
are observationally equivalent.

\begin{definition}
\label{def:obs}
Two states $\sigma_{1}$ and $\sigma_{2}$ are \textbf{\emph{observationally
	equivalent}}, written as $\sigma_{1} \thicksim \sigma_{2}$, if the return
	value of every operation supported by the data type applied on the two states
	is the same. Formally,
\begin{equation*}
	\begin{array}{l}
\forall \sigma_{1}, \sigma_{2} \in \Sigma.\ \forall o \in Op_{\tau}.\ \forall t_1,t_2 \in Timestamp.\ \exists a \in Val_{\tau}. \\
		\qquad \mathcal{D}_{\tau}.do (o,\sigma_{1},t_1) = (\_,a) \wedge \mathcal{D}_{\tau}.do (o,\sigma_{2},t_2) = (\_,a) \\
		\implies \sigma_{1} \thicksim \sigma_{2}
	\end{array}
\end{equation*}
\end{definition}

\begin{definition}
An execution $\chi$ (as in equation \ref{eq:2}) is \textbf{\emph{convergent modulo
	observable behavior}}, if for every state $(\phi_{i}, \delta_{i})$ and
\begin{equation}
\label{eq:convmod}
\forall b_{1}, b_{2} \in dom (\phi_i). \delta_{i} (b_{1}) = \delta_{i} (b_{2}) \implies \phi_{i} (b_{1}) \thicksim \phi_{i} (b_{2})
\end{equation}
\end{definition}

The idea behind convergence modulo observable behaviout is that the state of
the object at different replicas may not converge to the same (structurally
equal) representation, but the object has the same observable behaviour in
terms of its operations. For example, in the OR-set implementation, if the set
is implemented internally as a binary search tree (BST), then branches can
independently decide to perform balancing operations on the BST to improve the
complexity of the subsequent read operations. This would mean that the actual
state of the BSTs at different branches may eventually not be structurally
equal, but they would still contain the same set of elements, resulting in same
observable behaviour. Note that the standard notion of eventual consistency
implies convergence modulo observable behaviour.

\begin{definition}
\label{def:correct}
A data type implementation $\mathcal{D}_{\tau}$ is \textbf{\emph{correct}}, if
	every execution $\chi$ of $\mathcal{M}_{D_{\tau}}$ satisfies the
	specification $\mathcal{F}_\tau$ and is convergent modulo observable
	behavior.
\end{definition}

\section{Proving Data Type Implementations Correct}
\label{sec:correctness}

In the previous section, we have defined what it means for an MRDT
implementation to be correct with respect to the specification. In this
section, we show how to prove the correctness of an MRDT implementation with
the help of replication-aware simulation relations.

\subsection{Replication-aware simulation}

For proving the correctness of a data type implementation $\mathcal{D}_{\tau}$,
we use \textbf{\emph{replication-aware simulation relations}}
$~\mathcal{R}_{sim}$. While similar to the simulation relations used in Burckhardt
et al.~\cite{Burckhardt}, in this work, we apply them to MRDTs rather than
CRDTs. Further, we also mechanize and automate simulation-based proofs by
deriving simple sufficient conditions which can easily discharged by tools such
as F*. Finally, we apply our proof technique on a wide range of MRDTs, with
substantially complex specifications (e.g. queue MRDT described in
\S\ref{sec:queue}).

\begin{figure}[ht]
    \centering
    \includegraphics[scale=0.3]{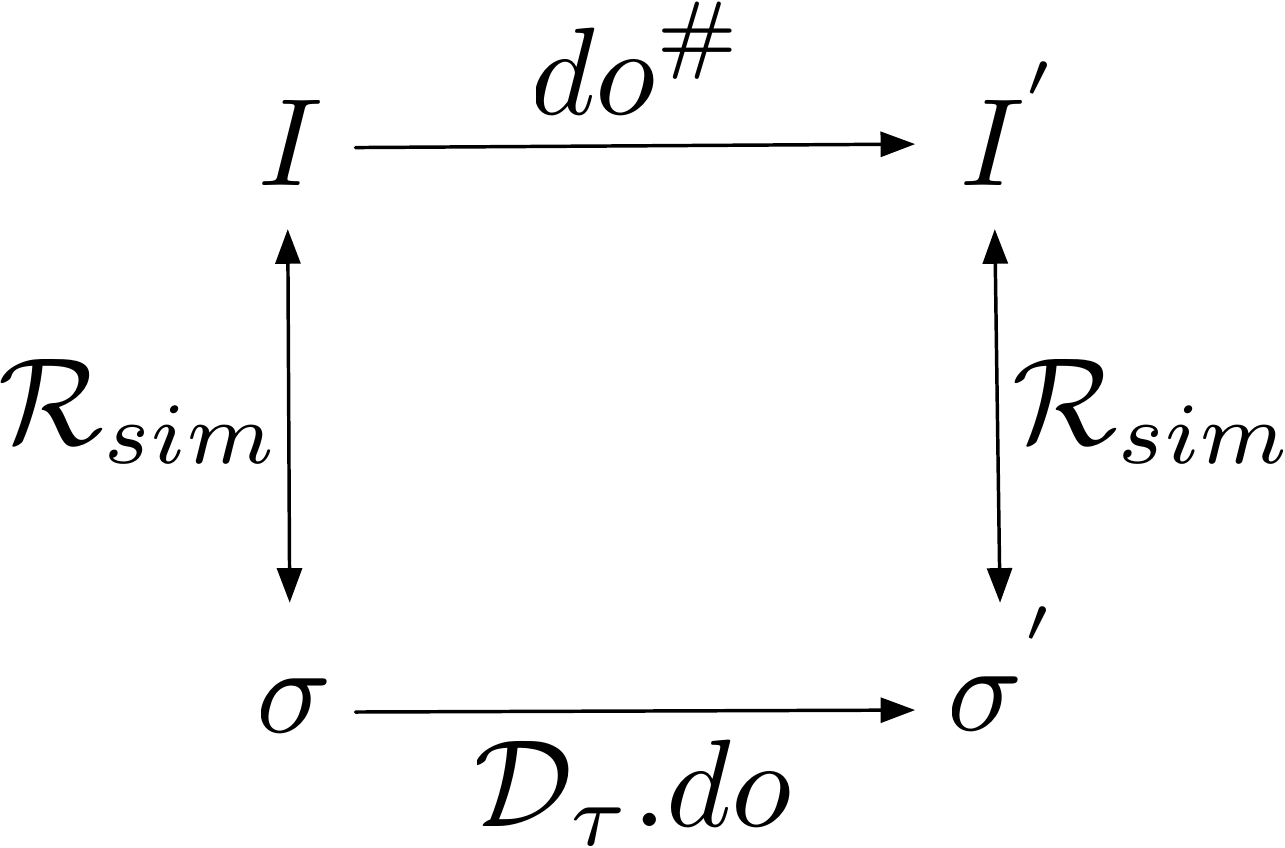}
		\caption{Verifying operations}
		\label{fig:fig4}
\end{figure}

\begin{figure}[ht]
    \centering
    \includegraphics[scale=0.3]{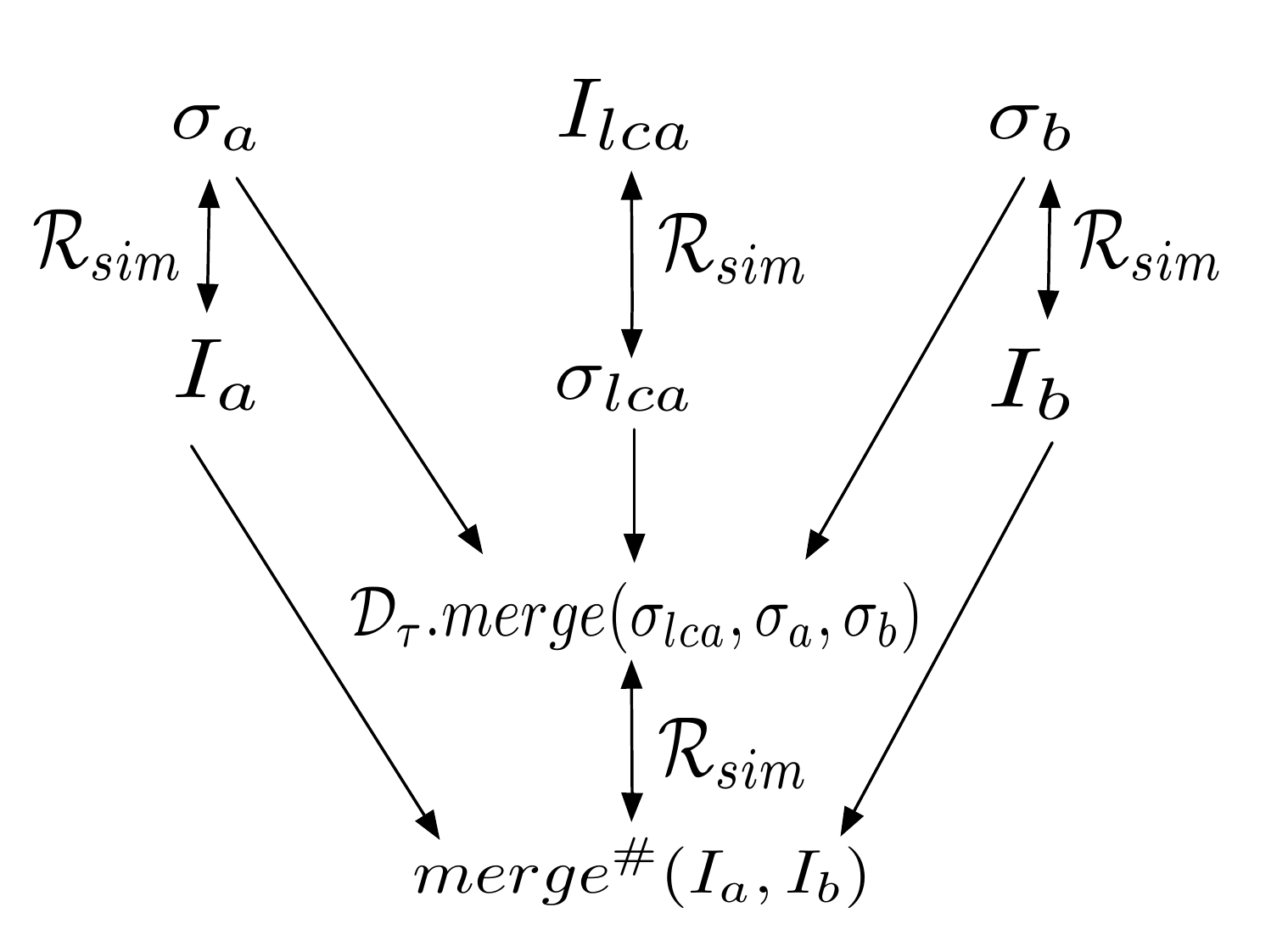}
		\caption{Verifying 3-way merge}
		\label{fig:fig5}
\end{figure}

The $\mathcal{R}_{sim}$ relation essentially associates the concrete state of
the object at a branch $b$ with the abstract state at the branch. This abstract
state would consist of all events which were applied on the branch. Verifying
the correctness of a MRDT through simulation relations involves two steps: (i)
first, we show that the simulation relation holds at every transition in every
execution of the replicated store, and (ii) the simulation relation meets the
requirements of the data type specification and is sufficient for convergence.
The first step is essentially an inductive argument, for which we require the
simulation relation between the abstract and concrete states to hold for every
data type operation instance and merge instance. These two steps are depicted
pictorially in figures \ref{fig:fig4} and \ref{fig:fig5}, respectively.

Figure \ref{fig:fig4} considers the application of a data type operation
(through the $do$ function) at a branch. Assuming that the simulation relation
$\mathcal{R}_{sim}$ holds between the abstract state $I$ and the concrete state
$\sigma$ at the branch, we would have to show that $\mathcal{R}_{sim}$
continues to hold after the application of the operation through the concrete
$do$ function of the implementation and the abstract $do^{\#}$ function on the
abstract state.

Figure~\ref{fig:fig5} considers the application of a merge operation between
branches $a$ and $b$. In this case, assuming $\mathcal{R}_{sim}$ between the
abstract and concrete states at the two branches and for the LCA, we would then
show that $\mathcal{R}_{sim}$ continues to hold between the concrete and
abstract states obtained after merge. Note that since the concrete merge
operation also uses the concrete LCA state $\sigma_{lca}$, we also assume that
$\mathcal{R}_{sim}$ holds between the concrete and abstract LCA states.

\begin{table*}[h]
\caption{\small Store properties}
\begin{tabular}{c | c}
\hline
$\Psi_{ts}(I)$\hspace{2pt}  & \hspace{2pt} $\forall e,e' \in I.E.\ e \xrightarrow{I.vis} e' \Rightarrow I.time(e) < I.time(e')  $ \\
 &$\wedge \forall e,e' \in I.E.\ I.time(e) = I.time(e') \Rightarrow e = e' $\\
\hline
$\Psi_{lca}(I_l, I_a, I_b)$ \hspace{2pt}  & \hspace{2pt} $I_l.vis = I_a.vis_{\mid I_l.E} = I_b.vis_{\mid I_l.E}$\\
& $ \wedge \forall e \in I_l.E.\ \forall e' \in (I_a.E \cup I_b.E) \setminus I_l.E.\ e \xrightarrow{I_a.vis \cup I_b.vis} e'$\\
\hline
\end{tabular}
\label{tab:global}
\end{table*}

These conditions consider the effect of concrete and abstract operations
locally and thus enable automated verification. In order to discharge these
conditions, we also consider two store properties, $\Psi_{ts}$ and
$\Psi_{lca}$ that hold across all executions (shown in Table \ref{tab:global}).
$\Psi_{ts}$ pertains to the nature of the timestamps associated with each
operation, while $\Psi_{lca}$ characterizes the lowest common ancestor used for
merge. These properties hold naturally due to the semantics of the replicated
store. These properties play an important role in
discharging the conditions required for validity of the simulation relation.

In particular, $\Psi_{ts}(I)$ asserts that in the abstract state $I$, causally
related events have increasing timestamps, and no two events have the same
timestamp. $\Psi_{lca}(I_l, I_a, I_b)$ will be instantiated with the LCA of two
abstract states $I_a$ and $I_b$ (i.e. $I_l = lca^{\#}(I_a, I_b)$), and asserts
that the visibility relation between events which are present in both $I_a$ and
$I_b$ (and hence also in $I_l$) will be the same in all three abstract states.
Further, every event in the LCA will be visible to newly added events in either
of the two branches. These properties follow naturally from the definition of
LCA and are also maintained by the store semantics.

Table \ref{tab:cond} shows the conditions required for proving the validity of
the simulation relation $\mathcal{R}_{sim}$. In particular, $\Phi_{do}$ and
$\Phi_{merge}$ exactly encode the scenarios depicted in the figures
\ref{fig:fig4} and \ref{fig:fig5}. Note that for $\Phi_{do}$, we assume
$\Psi_{ts}$ for the input abstract state on which the operation will be
performed. Similarly, for $\Phi_{merge}$, we assume $\Psi_{ts}$ for all events
in the merged abstract state (thus ensuring $\Psi_{ts}$ also holds for events
in the original branches) and $\Psi_{lca}$ for the LCA of the abstract states.

\begin{table*}[h]
\caption{Sufficient conditions for showing validity of simulation relation}
\begin{tabular}{c | c}
\hline
$\Phi_{do}(\mathcal{R}_{sim})$ & $\forall I,\sigma,e,op,a,t.\ \mathcal{R}_{sim} (I, \sigma) \wedge do^{\#} (I, e, op, a, t) = I^{'}$\\
& $\wedge~ \mathcal{D}_{\tau}.do (op, \sigma, t) = (\sigma^{'}, a) \wedge \Psi_{ts}(I)
	\implies \mathcal{R}_{sim} (I^{'}, \sigma^{'}) $\\
\hline
$\Phi_{merge}(\mathcal{R}_{sim})$ & $\forall I_a, I_b,\sigma_a,\sigma_b,\sigma_{lca}.\ \mathcal{R}_{sim}(I_a, \sigma_a) \wedge \mathcal{R}_{sim}(I_b, \sigma_b)$\\
& $\wedge~ \mathcal{R}_{sim}(lca^{\#}(I_a,I_b),\sigma_{lca}) \wedge \Psi_{ts}(merge^{\#}(I_a,I_b)) \wedge \Psi_{lca}(lca^{\#}(I_a,I_b), I_a, I_b)$\\
& $\implies \mathcal{R}_{sim}(merge^{\#}(I_a,I_b),\mathcal{D}_\tau.merge(\sigma_{lca}, \sigma_a, \sigma_b))$\\
\hline
$\Phi_{spec}(\mathcal{R}_{sim})$ & $\forall I,\sigma,e,op,a,t.\ \mathcal{R}_{sim} (I, \sigma) \wedge do^{\#} (I, e, op, a, t) = I^{'}$\\
& $\wedge~ \mathcal{D}_{\tau}.do (op, \sigma, t) = (\sigma^{'}, a) \wedge \Psi_{ts}(I)
	\implies a = \mathcal{F}_\tau(o,I)$\\
\hline
$\Phi_{con}(\mathcal{R}_{sim})$ & $\forall I,\sigma_a,\sigma_b.\ \mathcal{R}_{sim} (I, \sigma_{a}) \wedge \mathcal{R}_{sim} (I, \sigma_{b}) \implies \sigma_{a} \thicksim \sigma_{b}$\\
\hline
\end{tabular}
\label{tab:cond}
\end{table*}

Once we show that the simulation relation is maintained at every transition in
every execution inductively, we also have to show that it is strong enough to
imply the data type specification as well as guarantee convergence. For this,
we define two more conditions $\Phi_{spec}$ and $\Phi_{con}$ (also in table
\ref{tab:cond}). $\Phi_{spec}$ says that if abstract state $I$ and concrete
state $\sigma$ are related by $\mathcal{R}_{sim}$, then the return value of
operation $o$ performed on $\sigma$ should match the value of the specification
function $\mathcal{F}_\tau$ on the abstract state. Since the
$\mathcal{R}_{sim}$ relation is maintained at every transition, if
$\Phi_{spec}$ is valid, then the implementation will clearly satisfy the
specification. Finally, for convergence, we require that if two concrete states
are related to the same abstract state, then they should be observationally
equivalent. This corresponds to our proposed notion of convergence modulo observable behavior.

\begin{definition}
Given a MRDT implementation $\mathcal{D}_\tau$ of data type $\tau$, a
	replication-aware simulation relation $\mathcal{R}_{sim} \subseteq
	\mathcal{I}_\tau \times \Sigma$ is valid if $\Phi_{do}(\mathcal{R}_{sim})
	\wedge \Phi_{merge}(\mathcal{R}_{sim}) \wedge \allowbreak
	\Phi_{spec}(\mathcal{R}_{sim}) \allowbreak \wedge \allowbreak
	\Phi_{con}(\mathcal{R}_{sim})$.
\end{definition}

\begin{theorem} [Soundness]
\label{sound}
Given a MRDT implementation $\mathcal{D}_\tau$ of data type $\tau$, if there
	exists a valid replication-aware simulation $\mathcal{R}_{sim}$, then the
	data type implementation $\mathcal{D}_{\tau}$ is correct
	\footnote{The proof of the soundness theorem can be found in Appendix \ref{sec:proof}}.
\end{theorem}

\subsection{Verifying OR-sets using simulation relations}

Let us look at the simulation relations for verifying OR-set implementations in
\S\ref{sec:impl} against the specification $\mathcal{F}_{orset}$ in
\S\ref{sec:spec}.

\paragraph{OR-set.} Following is a candidate valid simulation relation for the
unoptimized OR-set from \S\ref{sec:or_set}:
\begin{equation}
\begin{split}
	\mathcal{R}_{sim} (I, \sigma)  \iff (\forall (a,t) \in \sigma \iff \\
		(\exists e \in I.E \wedge I. ~oper(e) = add(a) \wedge I.time(e) = t ~\wedge\\
		\neg (\exists f \in I.E \wedge I. ~oper(f) = remove(a) \wedge e \xrightarrow{vis} f)))
\end{split}
\end{equation}

The simulation relation says that for every pair of an element and a timestamp
in the concrete state, there should be an $\F{add}$ event in the abstract state
which adds the element with the same timestamp, and there should not be a
remove event of the same element which witnesses that add event. This
simulation relation is maintained by all the set operations as well as by the
merge operation, and it also matches the OR-set specification and guarantees
convergence. We use F* to automatically discharge all the proof obligations of
Table \ref{tab:cond}.

\paragraph{Space-efficient OR-set.} Following is a candidate valid simulation
relation for the space-efficient OR-set (OR-set-space) from
\S\ref{sec:or_set_space}:

\begin{equation}
\begin{split}
	\mathcal{R}_{sim} ((E, oper, rval, time, vis), \sigma)  \iff \\
	(\forall (a, t) \in \sigma \implies
			(\exists e \in E. ~oper(e) = add(a) \wedge time(e) = t \\
					\wedge~ \neg (\exists r \in E. ~oper(r) = remove(a) \wedge e \xrightarrow{vis} r)) ~\wedge \\
			(\forall e \in E. ~oper(e) = add(a) \wedge \neg (\exists r \in E. oper(r) = remove(a) \\
					\wedge~ e \xrightarrow{vis} r) \implies t \geq time(e))) ~\wedge \\
	(\forall e \in E. \forall a \in \mathbb{N}. ~oper(e) = add(a) \\
			\wedge~ \neg (\exists r \in E. ~oper(r) = remove(a) \wedge e \xrightarrow{vis} r) \implies (a, \_) \in \sigma)
\end{split}
\end{equation}

The simulation relation in this case captures all the constraints of the one
for OR-set with duplicates, but has additional constraints on the timestamp of
the elements in the concrete state. In particular, for an element in the
concrete state, the timestamp associated with that element will be the greatest
timestamp of all the $\F{add}$ events of the same element in the abstract
state, which has not been witnessed by a $\F{remove}$ event. Finally, we also
need to capture the constraint in the abstract to concrete direction. If there
is an $\F{add}$ event not seen by a $\F{remove}$ event on the same element,
then the element is a member of the concrete state. As before, the proof
obligations of Table \ref{tab:cond} are through F*.

\section{Composing MRDTs}
\label{sec:comp}

A key benefit of our technique is that compound data types can be constructed
by the composition of simpler data types through parametric polymorphism. The
proofs of correctness of the compound data types can be constructed from the
proofs of the underlying data types.

\subsection{IRC-style chat}
\label{subsec:chat}

To illustrate the benefits of compositionality, we consider a decentralised
IRC-like chat application with multiple channels. Each channel maintains the
list of messages in reverse chronological order so that the most recent message
may be displayed first. For simplicity, we assume that the channels are
append-only; while new messages can be posted to the channels, old messages
cannot be deleted. We also assume that while new channels may be created,
existing channels may not be deleted.

\begin{figure}[ht]
\begin{algorithmic} [1]
	\Statex $\mathcal{F}_{chat}(rd (ch), \langle E, oper, rval, time, vis \rangle) = log$ \text{where}
	\State $\quad (\forall t,m. ~(t,m) \in log \iff \exists e \in E. $
	\Statex $\quad \qquad \qquad \qquad  ~oper (e) = send (ch,m) \wedge~ time (e) = t) ~\wedge$
	\State $\quad (\forall t_1,m_1,t_2,m_2.~ ord ~(t_1, m_1) ~(t_2, m_2) ~log$
	\Statex $\quad \iff \exists e_1, e_2 \in E. ~oper (e_1) = send (ch, m_1) ~\wedge$
	\Statex $\quad \qquad \qquad time (e_1) = t_1 ~\wedge oper (e_2) = send (ch, m_2) ~\wedge$
	\Statex $\quad \qquad \qquad time (e_2) = t_2 \wedge t_1 > t_2)$
	\end{algorithmic}
\caption{The specification of IRC-style chat.}
\label{fig:chatexplicit}
\end{figure}

The chat application supports sending a message to a channel and reading
messages from a channel: $Op_{chat} = \{send (ch, m) \mid ch \in string \
~\wedge~ m \in string) \} ~\cup~ \{rd (ch) \allowbreak \mid ch \in string\}$.
The specification of this chat application is given in
Figure~\ref{fig:chatexplicit}. For this we define a predicate $ord$ such that
$ord (t_1, m_1) ~(t_2, m_2) ~l$ holds iff $t_1 \neq t_2$ and $(t_1, m_1)$
occurs before $(t_2, m_2)$ in list $l$. The specification essentially says the
log of messages contains all (and only those) messages that were sent, and
messages are ordered in reverse chronological order.

Rather than implement this chat application from scratch, we may quite
reasonably build it using existing MRDTs. We may use a MRDT \emph{map} to store
the association between the channel names and the list of messages. Given that
the conversations take place in a decentralized manner, the list of messages in
each channel should also be mergeable. For this purpose, we use a mergeable
\emph{log}, an MRDT list that totally orders the messages based on the message
timestamp, to store the messages in each of the channels. As mentioned earlier,
for simplicity we will assume that the map and the log are grow-only.

\subsection{Mergeable log}
\label{sec:mergelog}

\begin{figure}[ht]
\begin{algorithmic} [1]
	\Statex $\mathcal{F}_{log}(rd, \langle E, oper, rval, time, vis \rangle) = lst$ where
	\State $~\ (\forall t,m.~ (t,m) \in lst \iff$
	\Statex $~\ \qquad \exists e \in E.~oper(e) = append(m) ~\wedge~ time (e) = t) ~\wedge$
	\State $~\ (\forall t_1,m_1,t_2,m_2.~ ord ~(t_1,m_1) ~(t_2, m_2) ~lst \iff$
	\Statex $\quad \exists e_1, e_2 \in E. ~oper(e_1) = append(m_1) ~\wedge~ time(e_1) = t_1 $
	\Statex $\quad \wedge~ oper (e_2) = append (m_2) ~\wedge~ time (e_2) = t_2 ~\wedge~ t_1 > t_2)$
	\Statex
\end{algorithmic}

\begin{algorithmic} [1]
	\Statex $\mathcal{D}_{log} = (\Sigma, \sigma_{0}, do, merge_{log}) \text{ where }$
	\State $\quad \Sigma_{log} = \mathcal{P} (\mathbb{N} \times string)$
	\State $\quad \sigma_{0} = \{ \}$
	\State $\quad do(append (m), \sigma, t) = ((t, m) :: \sigma, \bot)$
	\State $\quad do(rd, \sigma, t) = (\sigma, \sigma)$
	\State $\quad merge_{log}(\sigma_{lca}, \sigma_a, \sigma_b) =$
	\Statex $\quad \qquad sort ((\sigma_{a} - \sigma_{lca}) ~@ ~(\sigma_{b} - \sigma_{lca})) ~@ ~\sigma_{lca} $
	\Statex
\end{algorithmic}

\begin{algorithmic} [1]
	\Statex $\mathcal{R}_{sim-log} (I, \sigma) \iff$
	\State $(\forall t,m.~ (t,m) \in \sigma \iff$
	\Statex $\quad \exists e \in I.E.~oper (e) = append (m) ~\wedge ~time (e)=t) ~\wedge$
	\State $\quad (\forall t_1,m_1,t_2,m_2. ~ord ~(t_1, m_1) ~(t_2, m_2) ~\sigma \iff $
	\Statex $\quad \exists e_1, e_2 \in I.E. ~oper (e_1) = append (m_1) ~\wedge~ time (e_1) = t_1$
	\Statex $\quad \wedge~ oper(e_2) = append (m_2) ~\wedge~ time (e_2) = t_2 ~\wedge~ t_1 > t_2)$
\end{algorithmic}
\caption{The specification, implementation and the simulation relation of mergeable log.}
\label{fig:log}
\end{figure}

The mergeable log MRDT supports operations to append messages to the log and to
read the log: $Op_{log} = \{rd\} \allowbreak ~\cup~ \allowbreak \{append(m) \mid m \in string\}$. The
log maintains messages in reverse chronological order. Figure~\ref{fig:log}
presents the specification, implementation and the simulation relation of the
mergeable log. The $sort$ function sorts the list in reverse chronological
order based on the timestamps associated with the messages.

\subsection{Generic map}
\label{subsec:alphamap}

\begin{figure}[ht]
\begin{algorithmic} [1]
	\Statex $\mathcal{F}_{\alpha-map}(get (k, o_{\alpha}), I) =$
	\Statex $\quad \text{let } I_{\alpha} = project (k, I) \text{ in } \mathcal{F}_{\alpha} (o_{\alpha}, I_{\alpha})$
	\Statex
\end{algorithmic}

\begin{algorithmic} [1]
	\Statex $\mathcal{D}_{\alpha-map} = (\Sigma, \sigma_{0}, do, merge_{\alpha-map}) \text{ where }$
	\State $\quad  \Sigma_{\alpha-map} = \mathcal{P} (string \times \Sigma_{\alpha})$
	\State $\quad \sigma_{0} = \{ \}$
	\State $\quad
   			 \delta (\sigma, k) =
   			 \begin{cases}
      				\sigma (k), & \text{if}\ k \in dom (\sigma) \\
      				\sigma_{0_{\alpha}}, & \text{otherwise}
   			 \end{cases}$
	\State $\quad do(set (k, o_{\alpha}), \sigma, t) = $
		 \Statex $\quad \quad \text{let } (v,r) = do_{\alpha} (o_{\alpha}, \delta (\sigma, k), t) \text{ in } (\sigma [k \mapsto v], r)$
	\State $\quad do(get (k, o_{\alpha}), \sigma, t) = $
		\Statex $\quad \quad \text{let } (\_,r) = do_{\alpha} (o_{\alpha}, \delta (\sigma, k), t)  \text { in } (\sigma, r)$
	\State $\quad merge_{\alpha-map}(\sigma_{lca}, \sigma_a, \sigma_b) = $
		\Statex $\quad  \quad \{ (k, v) \mid ( k \in dom(\sigma_{lca}) ~\cup~ dom(\sigma_{a}) ~\cup~ dom(\sigma_{b})) ~\wedge$
		\Statex $\quad  \quad \qquad \qquad v = merge_{\alpha} (\delta (\sigma_{lca}, k), \delta (\sigma_{a}, k), \delta (\sigma_{b}, k)) $
	\Statex
\end{algorithmic}

\begin{algorithmic} [1]
	\Statex $\mathcal{R}_{sim-\alpha-map} (I, \sigma) \iff \forall k.$
	\State $(k \in dom (\sigma) \iff \exists e \in I.E. ~oper (e) = set (k, \_)) ~\wedge $
	\State $ \quad\mathcal{R}_{sim-\alpha} ~(project (k, I), ~\delta (\sigma, k))$
\end{algorithmic}
\caption{The specification, implementation and simulation relation of $\alpha$-map.}
\label{fig:amap}
\end{figure}

We introduce a generic map MRDT, $\alpha$-map, which associates string keys
with a value, where the value stored in the map is itself an MRDT. This
$\alpha$-map is parameterised on an MRDT $\alpha$ and its implementation
$\mathcal{D}_\alpha$, and supports $get$ and $set$ operations: $Op_{\alpha-map}
= \{get (k, o_{\alpha}) \mid k \in string \wedge o_\alpha \in Op_\alpha\} \cup
\{set (k, o_{\alpha}) \mid k \in string \wedge o_\alpha \in Op_\alpha \}$,
where $Op_{\alpha}$ denotes the set of operations on the underlying value MRDT.

Figure~\ref{fig:amap} shows the specification, implementation and the
simulation relation of $\alpha$-map. The implementations for \textit{get} and
$set$ operations both fetch the current value associated with the key $k$ (and
the initial state of $\mathcal{D}_\alpha$ if the key is not present in the
map), and apply the given operation $o_{\alpha}$ from the implementation
$\mathcal{D}_\alpha$ on this value. While $set$ updates the binding in the map,
$get$ does not do so and simply returns the value returned by $o_\alpha$. The
merge operation merges the values for each key using the merge function of
$\alpha$. The specification and simulation relation of $\alpha$-map use the
specification and simulation relation of the underlying MRDT $\alpha$, by
projecting the events associated with each key to an abstract execution of
$\alpha$. We now provide the details of this projection function.

\subsection{Projection function}
\label{subsec:proj}

\begin{figure}[ht]
\begin{algorithmic} [1]
	\Statex $project ~k ~I_{\alpha-map} = I_{\alpha} \text{ where }$
	\State $\quad I_{\alpha-map} = (\Sigma_m, oper_m, rval_m, time_m, vis_m)$ and
	\State $\quad I_{\alpha} = (\Sigma_\alpha, oper_\alpha, rval_\alpha, time_\alpha, vis_\alpha)$ and
	\State \quad $(\forall e,k,o.~ e \in \Sigma_m \wedge ~oper_m (e) = set (k, o) \iff$
	\Statex $\quad \exists e' \in \Sigma_\alpha. ~oper_\alpha (e') = o ~\wedge rval_m(e) = rval_\alpha(e') ~\wedge $
	\Statex $\quad \quad  ~ time_m(e) = time_\alpha(e')) ~\wedge$
	\State $\quad (\forall e_1, e_2.~ e_1 \in \Sigma_m \wedge e_2 \in \Sigma_m \wedge  ~oper_m (e_1) = set (k, \_) ~\wedge $
	\Statex $\quad oper_m (e_2) = set (k, \_)  ~\wedge e_1 \xrightarrow{vis_m} e_2 \iff $
	\Statex $\quad \exists e_{1}^{'}, e_{2}^{'} \in \Sigma_\alpha. ~time_\alpha (e_{1}^{'}) = time_m(e_1) ~\wedge $
	\Statex $\quad \quad ~time_\alpha (e_{2}^{'}) = time_m(e_2) ~\wedge e_{1}^{'} \xrightarrow{vis_\alpha} e_{2}^{'}) $
	\end{algorithmic}
\caption{Projection function for mapping $\alpha$-map execution to $\alpha$ execution.}
\label{fig:proj}
\end{figure}

Figure~\ref{fig:proj} gives the projection function which when given an abstract
execution $I_{\alpha-map}$ of $\alpha$-map, projects all the $set$-events
associated with a particular key $k$ to define an abstract execution
$I_\alpha$. There is a one-to-one correspondence between $set$-events to $k$ in
$I_{\alpha-map}$ and events in $I_\alpha$, with the corresponding events in
$I_\alpha$ preserving the operation type, return values, timestamps and the
visibility relation. The project function as used in the specification of
$\mathcal{F}_{\alpha-map}$ ensures that the return value of $get$-events obey
the specification $\mathcal{F}_\alpha$ as applied to the projected
$\alpha$-execution.

Similarly, the simulation relation of $\alpha$-map requires the simulation
relation of $\alpha$ to hold for every key, between the value associated with
the key and the corresponding projected execution for the key. We can now
verify the correctness of the generic $\alpha$-map MRDT by relying on the
correctness of $\alpha$. That is, if $\mathcal{R}_{sim-\alpha}$ is a valid
simulation relation for the implementation $\mathcal{D}_\alpha$, then
$\mathcal{R}_{sim-\alpha-map}$ is a valid simulation relation for
$\mathcal{D}_{\alpha-map}$. This allows us to build the proof of correctness of
$\alpha$-map using the proof of correctness of $\alpha$.

\begin{figure}[ht]
\begin{algorithmic} [1]
	\Statex $\mathcal{F}_{chat}(rd(ch),I) = \mathcal{F}_{log-map}(get(ch,rd),I)$
	\Statex
	\Statex $\mathcal{D}_{chat} =  \mathcal{D}_{log-map} \text{ where }$
	\State $\qquad do(send(ch,m),\sigma,t) =$
	\Statex $\qquad \qquad do(set(ch,append(m)),\sigma,t)$
	\State $\qquad do(rd(m), \sigma, t) = do(get(k,rd), \sigma, t)$
\end{algorithmic}
\caption{Implementation of IRC-style chat.}
\label{fig:chat}
\end{figure}

For our chat application, we instantiate $\alpha$-map with the mergeable log
$\mathcal{D}_{log}$. The chat application itself is a wrapper around the
log-map MRDT as shown in Fig.~\ref{fig:chat}. In order to verify the
correctness of $\mathcal{D}_{chat}$, we only need to separetely verify
$\mathcal{D}_{\alpha-map}$ and $\mathcal{D}_{log}$. Note that one can
instantiate $\alpha$ with any verified MRDT implementation to obtained a
verified $\alpha$-map MRDT.

\section{Case study: A Verified Queue MRDT}
\label{sec:queue}

Okasaki~\cite{okasaki_1998} describes a purely functional queue with amortized
time complexity of $O(1)$ for enqueue and dequeue operations. This queue is
made up of two lists that hold the front and rear parts of the queue. Elements
are enqueued to the rear queue and dequeued from the front queue (both are $O(1)$
operations). If the front queue is found to be empty at dequeue, then the rear
queue is reversed and made to be the front queue ($O(n)$ operation). Since each
element is part of exactly one reverse operation, the enqueue and the dequeue
have an amortized time complexity of $O(1)$. In this section, we show how to
convert this efficient sequential queue into an MRDT by providing
additional semantics to handle concurrent operations.

For simplicity of specification, we tag each enqueued element with the unique
timestamp of the enqueue operation, which ensures that all the elements in the
queue are unique. The queue supports two operations: $Op_{queue} = \{dequeue\}
\cup \{enqueue(a) \mid a \in \mathbb{V} \}$, where $\mathbb{V}$ is some value
domain. Unlike a sequential queue, we follow an \emph{at-least-once} dequeue
semantics -- an element inserted into the queue may be consumed by concurrent
dequeues on different branches. At-least-once semantics is common for
distributed queueing services such as Amazon Simple Queue
Service(SQS)~\cite{SQS_Amazon} and RabbitMQ~\cite{Q_Rabbitmq}. At a merge,
concurrent enqueues are ordered according to their timestamps.

\subsection{Merge function of the replicated queue}

\begin{figure}[ht]
	\centering
	\includegraphics[scale=0.5]{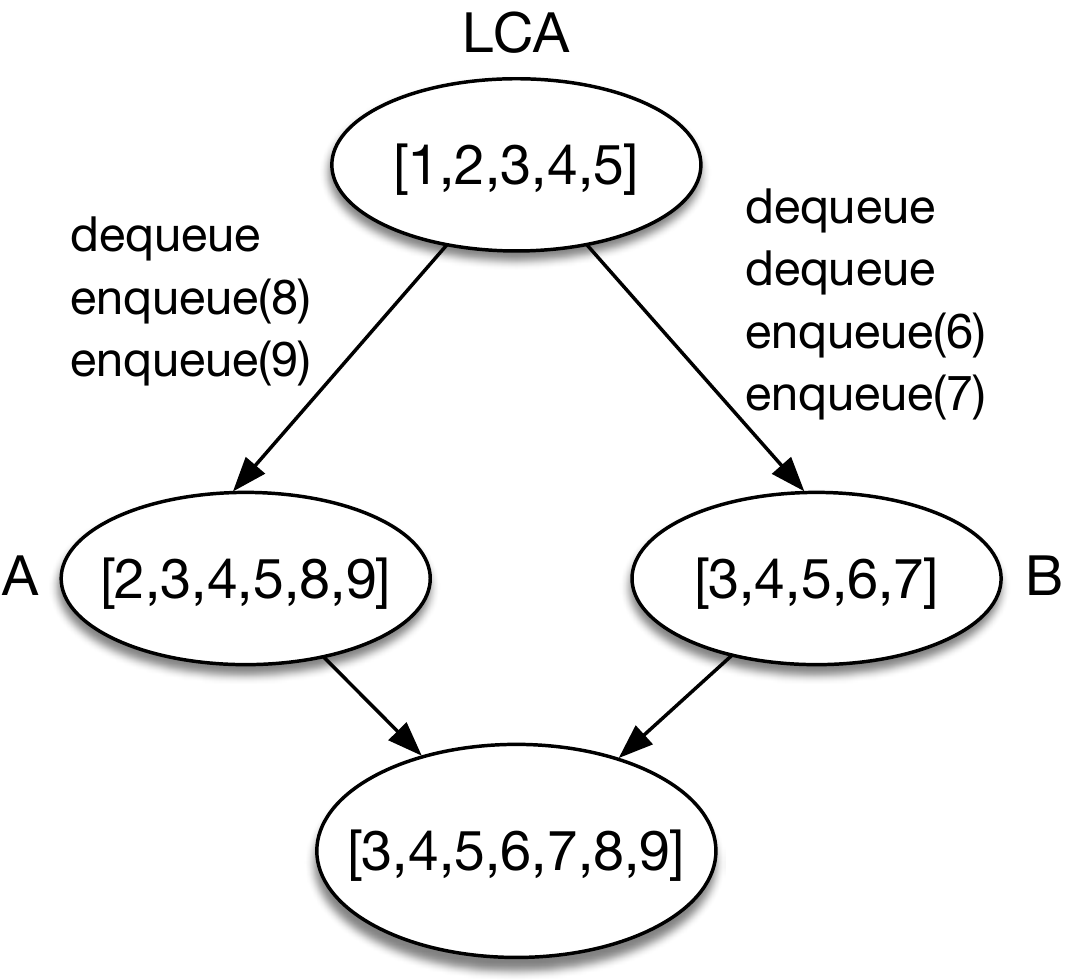}
	\caption{Three-way merge for queues}
	\label{fig:fig10}
\end{figure}

To illustrate the three-way merge function, consider the execution presented in
figure~\ref{fig:fig10}. For simplicity, we assume that the timestamps are the
same as the values enqueued. Starting from the LCA, each branch performs a
sequence of dequeue and enqueue operations. The resulting versions are then
merged. Observe that in the merged result, the elements 1 and 2 which were
dequeued (with 1 dequeued on both the branches!) are not present. Elements 3, 4
and 5 which are present in all three versions are present in the merged result.
Newly inserted elements appear at the suffix, sorted according to their
timestamps.

The merge function first converts each of the queues to a list, and finds the
longest common contiguous subsequence between the three versions
(\emph{[3,4,5]}). The newly enqueued elements are suffixes of this common
subsequence -- \emph{[8,9]} and \emph{[6,7]} in the queues A and B,
respectively. The final merged result is obtained by appending the common
subsequence to the suffixes merged according to their timestamps. Each of these
operations has a time complexity of $O(n)$ where $n$ is the length of the
longest list. Hence, the merge function is also an $O(n)$ operation
\footnote{The implementation of the queue operations and the merge function is available in Appendix \ref{sec:queue1}}.

\subsection{Specification of the replicated queue}

We now provide the specification for the queue MRDT, which is based on the
declarative queue specification in Kartik et al.~\cite{Kartik1}. In particular,
compared to the sequential queue, the only constraint that we relax is allowing
multiple dequeues of the same element.

In order to describe the specification, we first introduce a number of axioms
which declaratively specify different aspects of queue behaviour. Consider the
$\F{match}_I$ predicate defined for a pair of events $e_1,e_2$ in an abstract
execution $I$:

\begin{align*}
 \F{match}_I(e_1,e_2) & \Leftrightarrow I.oper(e_1) = enqueue(a) \\
											& ~~\wedge I.oper(e_2) = dequeue  \wedge a = I.rval(e_2)
\end{align*}

\newcommand{\emptyv}{\mathtt{EMPTY}}

Let $\emptyv$  be the value returned by a dequeue when the queue is empty. We
define the following axioms:

\begin{itemize}
	\item $AddRem(I)$ : $\forall e \in I.E.\ I.oper(e) = dequeue ~\wedge $ \\
											$ I.rval(e) \neq \emptyv \implies \exists e' \in I.E.\ \F{match}_I(e', e) $

	\item $Empty(I)$ : $ \forall e_1,e_2, e_3 \in I.E.\ I.oper(e_1) = dequeue ~\wedge$ \\
		                 $ I.rval(e_1) = \emptyv ~\wedge~ I.oper(e_2) = enqueue(a) ~\wedge$ \\
										 $ e_2 \xrightarrow{I.vis} e_1 \implies \exists e_3 \in I.E.\ \F{match}_I(e_2,e_3) \wedge e_3 \xrightarrow{I.vis} e_1  $

	\item $FIFO_1(I)$ : $ \forall e_1,e_2,e_3 \in I.E.\ I.oper(e_1) = enqueue(a) ~\wedge $ \\
											$ \F{match}_I(e_2,e_3) ~\wedge~ e_1 \xrightarrow{I.vis}e_2 \implies \exists e_4 \in I.E.\ \F{match}_I(e_1,e_4) $

	\item $FIFO_2(I)$: $ \forall e_1,e_2,e_3,e_4 \in I.E.\ \neg (\F{match}_I(e_1,e_4) ~\wedge $ \\
										 $ \F{match}_I(e_2,e_3) ~\wedge~ e_1 \xrightarrow{I.vis} e_2 ~\wedge~ e_3  \xrightarrow{I.vis}e_4) $
\end{itemize}

These axioms essentially encode queue semantics. $AddRem$ says that for every
dequeue event which does not return $\emptyv$, there must exist a matching
enqueue event. $Empty$ says that if a dequeue event returns $\emptyv$, there
should not be an unmatched enqueue visible to it. Finally, $FIFO_1$ and
$FIFO_2$ encode the first-in-first-out nature of the queue. These axioms ensure
that if an enqueue event $e_1$ was visible to another enqueue event $e_2$, then
the element inserted by $e_1$ will be dequeued first. Notice that sequential
queue would also have an injectivity axiom, which disallows multiple dequeues
to be matched to an enqueue, but we do not enforce this requirement for the
replicated queue.

To define $\mathcal{F}_{Queue}$, we first note that enqueue operation always
returns $\bot$. For an abstract state $I$, $\mathcal{F}_{Queue}(dequeue, I)$
returns $a$ such that if we add the new event $e$ for the dequeue to the
abstract state $I$, then the resulting abstract state
$do^{\#}(I,e,dequeue,a,t)$ must satisfy all the queue axioms.

Notice how the queue axioms are substantially different from the way the MRDT
queue is actually implemented. The simulation relation that we use to bridge
this gap and relate the implementation with the abstract state is actually very
straightforward: we simply say that for every element present in the concrete
state of the queue, there must be an enqueue event without a matching dequeue.
We also assert the other direction, and enforce the queue axioms on the
abstract state. The complete simulation relation can be found in the
supplemental material. We were able to successfully discharge the conditions
for validity of the simulation relation using F*.

\section{Evaluation}
\label{sec:results}

In this section, we evaluate the instantiation of the formalism developed thus
far in \name, an F* library of certified efficient MRDTs. We first discuss the
verification effort followed by the performance evaluation of efficient MRDTs
compared to existing work. These results were obtained on a 2-socket
Intel\textregistered Xeon\textregistered Gold 5120
x86-64~\cite{IntelXeonGold5120Spec} server running Ubuntu 18.04 with 64GB of
main memory.

\subsection{Verification in F*}

F*'s core is a functional programming language inspired by ML, with support for
program verification using refinement types and monadic effects. Though F* has
support for built-in effects, \name library only uses the pure fragment of the
language. Given that we can extract OCaml code from our verified
implementations in F*, we are able to directly utilise our MRDTs on top of
Irmin~\cite{Irmin}, a Git-like distributed database, whose execution model fits
the MRDT system model.

As part of the \name library, we have implemented and verified 9 MRDTs --
increment-only counter, PN counter, enable-wins flag, last-writer-wins
register, grows-only set, grows-only map, mergeable log, observed-remove set
and functional queue. Our specifications capture both the functional
correctness of local operations as well as the semantics of the concurrent
conflicting operations.

F*'s support for type classes provides a modular way to implement and verify
MRDTs. The \name library defines a MRDT type class that captures the sufficient
conditions to be proved for each MRDT as given in Table~\ref{tab:cond}. This
library contains 124 lines of F* code. Each MRDT is a specific instance of the
type class which satisfy the conditions. It is useful to note that our MRDTs
sets, maps and queues are polymorphic in their contents and may be plugged with
other MRDTs to construct more complex MRDTs as seen in \S\ref{sec:comp}.

\begin{table*}[ht]
	\caption{\name verification effort.}
	\label{table:MRDT}
	\begin{tabular}{p{0.3\linewidth} p{0.13\linewidth} p{0.13\linewidth} p{0.1\linewidth} p{0.15\linewidth}}
		\toprule
		\textbf{MRDTs verified}  &\textbf{\#Lines code}  &\textbf{\#Lines proof} &\textbf{\#Lemmas} &\textbf{Verif. time (s)} \\
		\midrule
		Increment-only counter &6 &43 &2 &3.494\\
		\midrule
		PN counter &8 &43 &2 &23.211\\
		\midrule
		Enable-wins flag &20 &58 &3 &1074\\
		& &81 &6 &171 \\
		& &89 &7 &104\\
		\midrule
		LWW register &5 &44 &1 &4.21\\
		\midrule
		G-set &10 &23 &0 &4.71\\
		& &28 &1 &2.462\\
		& &33 &2 &1.993\\
		\midrule
		G-map &48 &26 &0 &26.089\\
		\midrule
		Mergeable log &39 &95 &2 &36.562\\
		\midrule
		OR-set (\S\ref{sec:or_set})  &30 &36 &0 &43.85\\
		& &41 &1 &21.656\\
		& &46 &2 &8.829\\
		\midrule
		OR-set-space (\S\ref{sec:or_set_space})
		&59 &108 &7 &1716\\
		\midrule
		OR-set-spacetime &97 &266 &7 &1854\\
		\midrule
		Queue &32 &1123 &75 &4753\\
		\bottomrule
	\end{tabular}
\end{table*}

Table~\ref{table:MRDT} tabulates the verification effort for each MRDT in the \name
library. We include three versions of OR-sets:

\begin{itemize}
	\item \textbf{OR-set:} the unoptimized one from \S\ref{sec:or_set} which uses
		a list for storing the elements and contains duplicates.
	\item \textbf{OR-set-space:} the space-optimized one from
		\S\ref{sec:or_set_space} which also uses a list but does not have
		duplicates.
	\item \textbf{OR-set-spacetime:} a space- and time-optimized one which uses a
		binary search tree for storing the elements and has no duplicates. The
		merge function produces a height balanced binary tree.
\end{itemize}

The lines of code represents the number of lines for implementing the data
structure without counting the lines for refinements, lemmas, theorems and
proofs. This is approximately the number of lines of code there will be if the
data structures were implemented in OCaml. Everything else that has to do with
verification is included in the lines of proofs. It is useful to note that the
lines of proof for simple MRDTs such as counter and last-writer-wins (LWW)
register is high compared to the lines of code since we also specify and prove
their full functional correctness.

For many of the proofs, F* is able to automatically verify the properties
either without any lemmas or just a few, thanks to F* discharging the proof
obligations to the SMT solver. Most of the proofs are a few tens of lines of
code with the exception of queues. In queues, the implementation is far removed
from the specification, and hence, additional lemmas were necessary to bridge
this gap.

F* allows the user to provide additional lemmas that help the solver arrive to
the proof faster. We illustrate this for enable-wins flag, G-set and OR-set by
adding additional lemmas. Correspondingly, we observe that the verification
time reduces significantly. Thanks to F*, the developers of new MRDTs in \name
can strike a balance between verification times and manual verification effort.

In this work, we have not used F* support for tactics and interactive proofs.
We believe that some of the time consuming calls to the SMT solver may be
profitably replaced by a few interactive proofs. On the whole, the choice of F*
for \name reduces manual effort and most of the proofs are checked within few
seconds.

\subsection{Performance evaluation}

In this section, we evaluate the runtime performance of efficient MRDTs in
\name.

\subsubsection{\name vs Quark}
\label{sec:quark_vs_peepul}

We first compare the performance of \name MRDTs against the MRDTs presented in
Kaki et al.~\cite{Kaki} (Quark). Recall that Quark lifts sequential data types
to MRDTs by equipping them with a merge function, which converts the concrete
state of the MRDT to a relational (set-based) representation that captures the
characteristic relations of the data type. The actual merge is implemented as a
merge of these sets for each of the characteristic relations. After merge of
the relational representations, the final result is obtained by a
concretization function. Compared to this, \name merges are implemented
directly on the concrete representations.

To highlight the impact of the efficient merge function in \name, we evaluate
the performance of merge in queues. Both \name and Quark uses the same
sequential queue representation, and the only difference is the merge function
between the two. For this experiment, we start with an empty queue, and perform
a series of randomly generated operations with 75:25 split between enqueues and
dequeues. We use this version as the LCA and subsequently perform two different
sets of operations to get the two divergent versions. We then merge these
versions to measure the time taken for the merge.

\begin{figure}[ht]
	\centering
	\includegraphics[scale=0.4]{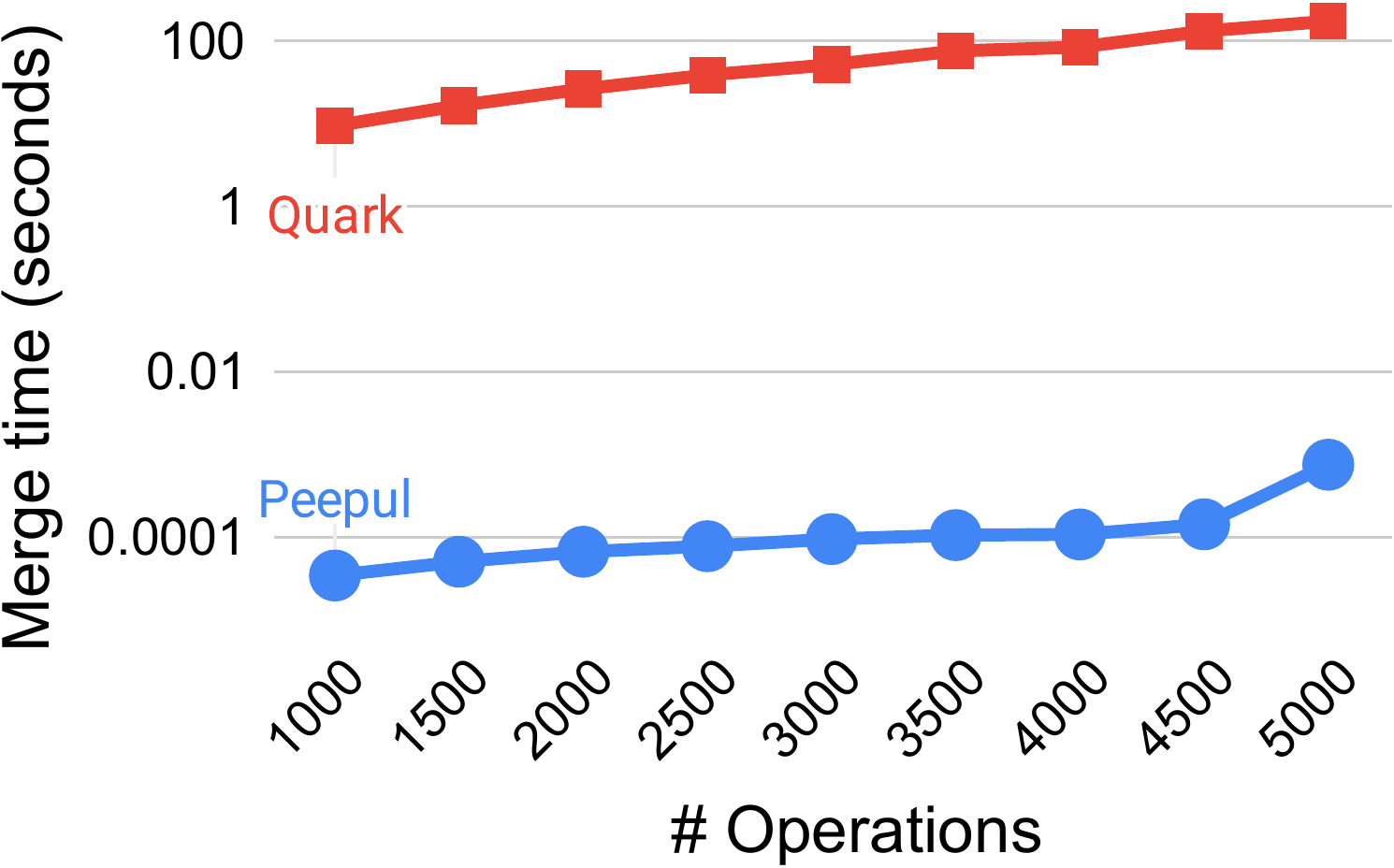}
	\caption{Merge performance of \name and Quark queues.}
	\label{fig:fqueue}
\end{figure}

The results are reported in figure~\ref{fig:fqueue}. For a queue, Quark needs
to reify the ordering relation as a set which will contain $n^2$ elements for a
queue of size $n$. In addition, there is also the cost of abstracting and
concretising the queue to and from relational representation. As a result, the
merge function takes 10 seconds for 1000 operations, increasing to 178 seconds
for 5000 operations. On the other hand, \name's linear-time merge took less
than a millisecond in all of the cases. This shows the that Quark merge is
unacceptably slow even reasonably sized queues, while \name remains fast and
practical.

\begin{figure}[ht]
	\centering
	\includegraphics[scale=0.4]{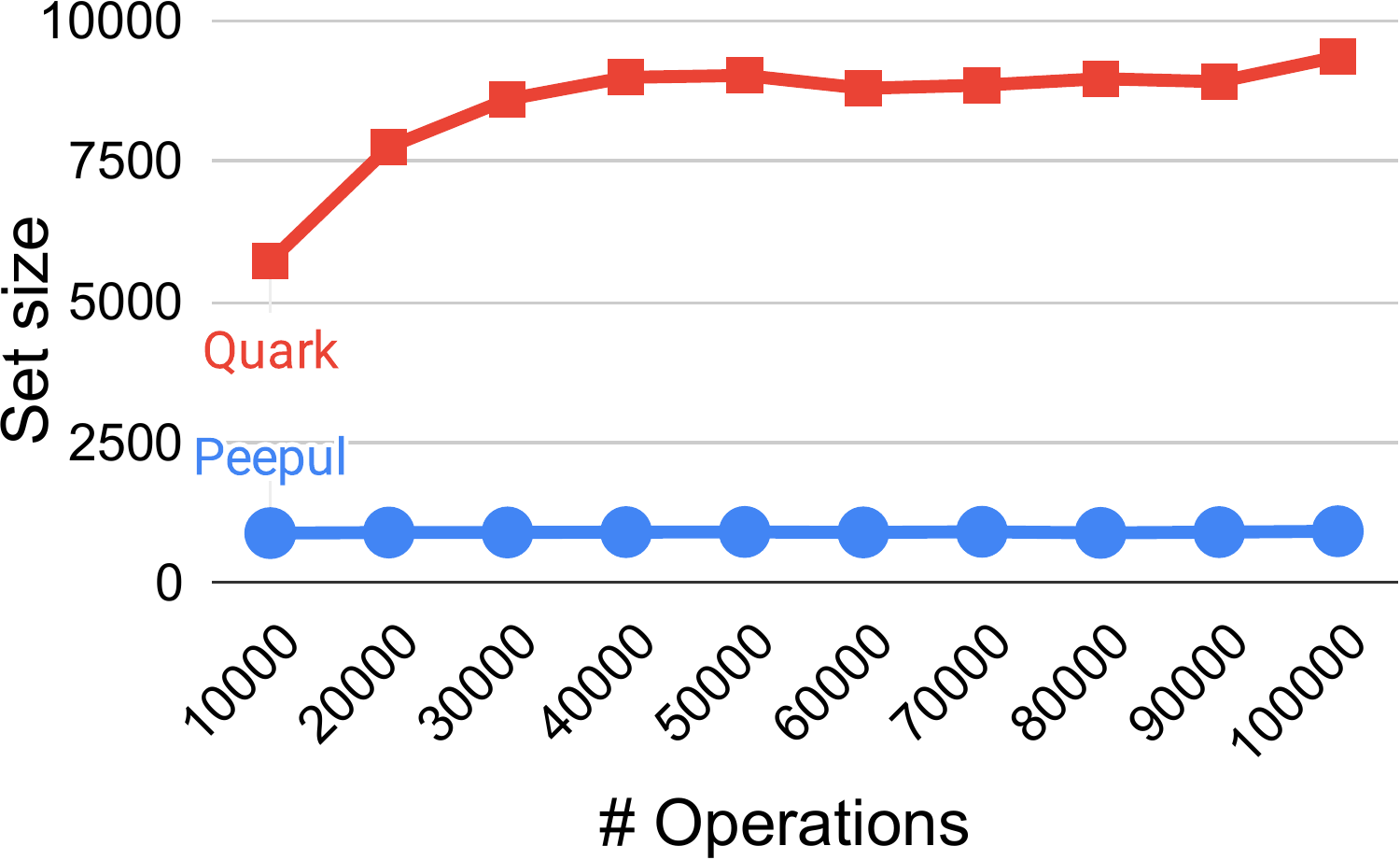}
	\caption{Performance of \name and Quark OR-sets.}
	\label{fig:orset}
\end{figure}

We also compare the performance of OR-set in \name and Quark. Since the merge
function in Quark is based on automatic relational reification, Quark does not
allow duplicate elements to be removed from the OR-set. To highlight the impact
of duplicate elements, we perform an experiment similar to the queue one except
that we pick a 50:50 split between add and remove operations. The values added
are randomly picked in the range (0:1000). For \name, we pick the
space-optimized OR-set (OR-set-space). We report the number of elements in the
final set including duplicates.

The results are presented in figure~\ref{fig:orset}. Due to the duplicates, the
size of the Quark set increases with increasing number of operations; the
growth is not linear due to the stochastic interplay between add and remove.
For \name, the set size always remains below 1000 which is the range of the
values picked. The results show that MRDTs in \name are much more efficient
than in Quark.

\subsubsection{\name OR-set performance}

\begin{figure}
	\centering
	\includegraphics[scale=0.4]{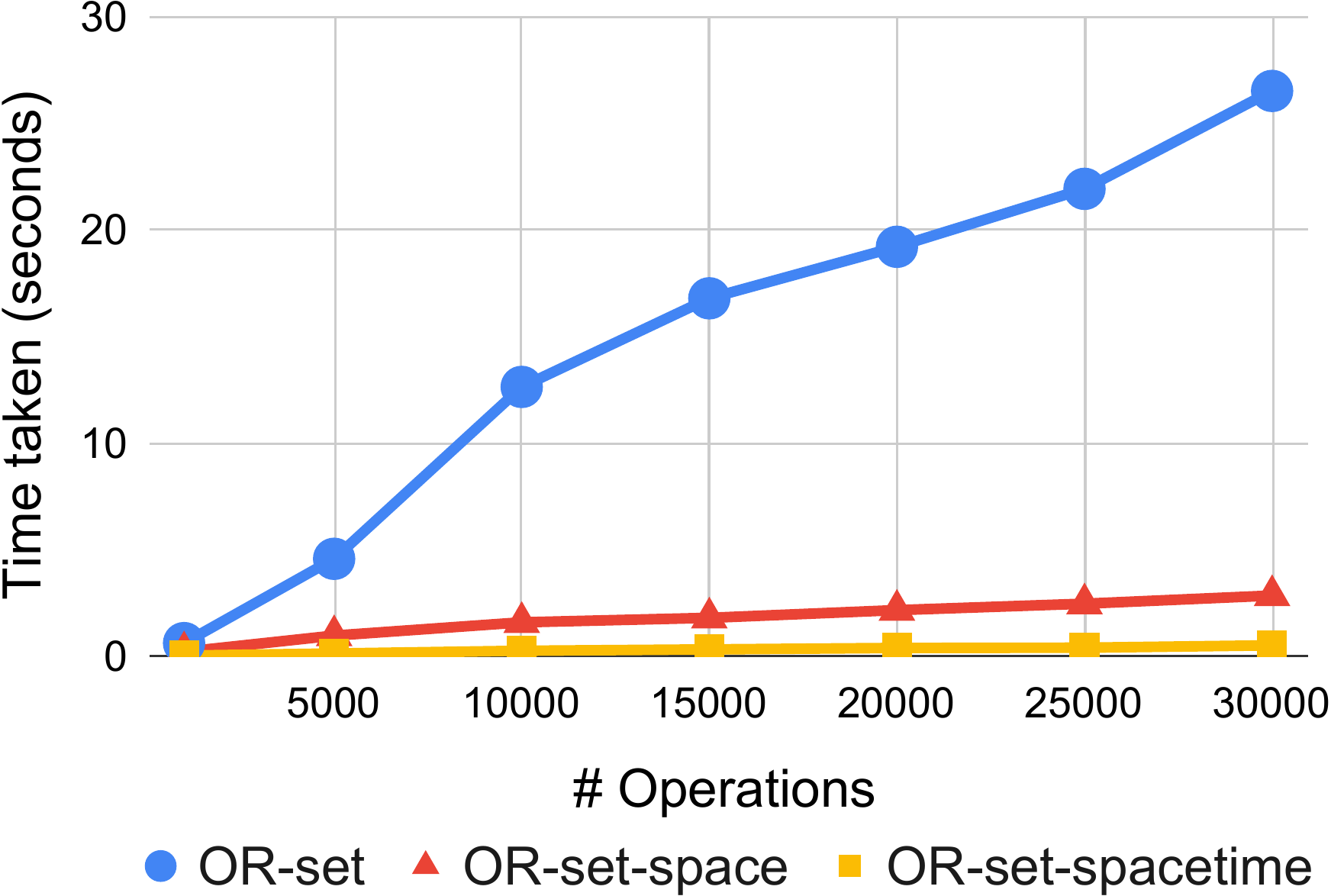}
	\caption{Running time of OR-sets.}
	\label{fig:orset1}
\end{figure}

\begin{figure}
	\centering
	\includegraphics[scale=0.4]{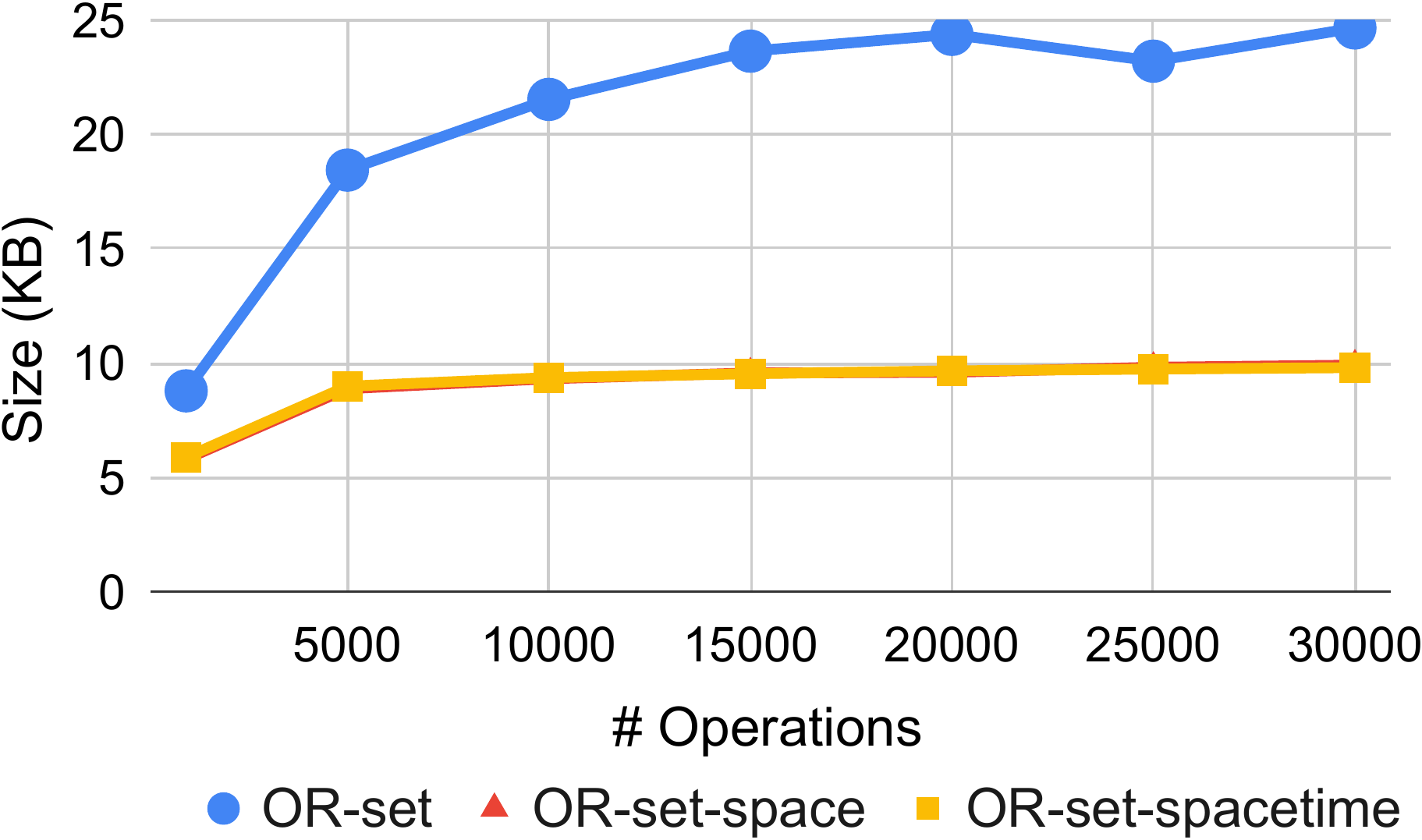}
	\caption{Space consumption of OR-sets. The OR-set-space line is hidden by the
	OR-set-spacetime line.}
	\label{fig:orset2}
\end{figure}

We also compare the overall performance of the three OR-set implementations in
\name. Our workload consists of 70\% lookups, 20\% adds and 10\% remove
operations starting from an initial empty set on two different branches. We
trigger a merge every 500 operations. We measure the overall execution time for
the entire workload and the maximum size of the set during the execution.

The results are reported in figures~\ref{fig:orset1} and \ref{fig:orset2}. The
results show that OR-set-spacetime is the fastest, and is around 5$\times$
faster than OR-set-space due to the fast reads and writes thanks to the binary
search tree in OR-set-spacetime. Both OR-set-space and OR-set-spacetime consume
similar amount of memory. The unoptimized OR-set is both slower and consumes
more memory than the other variants due to the duplicates. The results show
that \name enables construction of efficient certified MRDTs that have
significant performance benefits compared to unoptimised ones.

\section{Related Work}
\label{sec:related}

Reconciling concurrent updates is an important problem in distributed systems.
Some of the works proposing new designs and implementations of
RDTs~\cite{Shapiro, Roh, Annette} neither provide their formal specification
nor verify them. Due to the concurrently evolving state of the replicas,
informally reasoning about the correctness of even simple RDTs is tedious and
error prone. In this work, our focus is on mechanically verifying efficient
implementations of MRDTs.

There are several works that focus on specification and verification of
CRDTs~\cite{Burckhardt,Liu,Zeller,Nair,Attiya,Gomes,Kartik}. CRDTs typically
assume a system model which involves several replicas communicating over
network with asynchronous message passing. Correspondingly, the specification
and verification techniques for CRDTs will have to take into account of the
properties of message passing such as message ordering and delivery guarantees.
On the other hand, MRDTs are described over a Git-like distributed store with
branching and merging, which in turn may be implemented over asynchronous
message passing. We believe that, by lifting the level of abstraction, MRDTs
are easy to specify, implement and verify compared to CRDTs.

In terms of mechanised verification of RDTs, prior work has used both automated
and interactive verification. Zeller et al.~\cite{Zeller} verify state-based
CRDTs with the help of interactive theorem prover Isabelle/HOL. Gomes et
al.~\cite{Gomes} develop a foundational framework for proving the correctness
of operation-based CRDTs. In particular, they construct a formal network model
that may delay, drop or reorder messages sent between replicas. Under these
assumptions, they verify several op-based CRDTs using Isabelle/HOL. Nair et
al.~\cite{Nair} presents an SMT-based verification tool to specify state-based
CRDTs and verify invariants over its state. Kartik et al.~\cite{Kartik} also
utilise SMT-solver to automatically verify the convergence of CRDTs under
different weak consistency policies. Liu et al.~\cite{Liu} present an extension
of the SMT-solver-aided Liquid Haskell to allow refinement types on type
classes and use to implement a framework to verify operation-based CRDTs.
Similar to Liu et al., \name also uses an SMT-solver-aided programming language
F*. We find that SMT-solver-aided programming language offers a useful trade
off between manual verification effort and verification time.

Our verification framework for MRDTs builds on the concept of replication-aware
simulation introduced by Burckhardt et al.~\cite{Burckhardt}. Burckhardt et al.
present precise specifications for RDTs and (non-mechanized) proof of
correctness for a few CRDT implementations. Burckhardt et al.'s specifications
are presented over the CRDT system model with explicit message passing between
replicas. In this work, we lift these specifications to a higher level by
abstracting out the guarantees provided by the low-level store ($\Psi_{ts}$ and
$\Psi_{lca}$). Further, we also observe that the simulation relation
$\mathcal{R}_{sim}$ cannot be used as an inductive invariant on its own, and
instead, a conjunction of $\mathcal{R}_{sim}$ with $\Psi_{ts}$ and $\Psi_{lca}$
is required (see conditions $\Phi_{do}$ and $\Phi_{merge}$ in Table
\ref{tab:cond}). In order to enable mechanised verification, we identify the
relationship between $\mathcal{R}_{sim}$ and the functional correctness and
convergence of MRDTs. This leads to a formal specification framework that is
suitable for mechanized and automated verification. We demonstrate this by
successfully verifying a number of complex MRDT implementations in F* including
the first, formally verified replicated queue.

MRDTs were first introduced by Farnier et al.~\cite{farinier15} for
Irmin~\cite{Irmin}, a distributed database built on the principles of Git.
Quark~\cite{Kaki} automatically derives merge functions for MRDTs using
invertible relational specification. However, their merge semantics focused
only on convergence, and not the functional correctness of the data type. Our
evaluation (\S\ref{sec:quark_vs_peepul}) shows that merges through
automatically derived invertible relational specification is prohibitively
expensive for data types with rich structure such as queues.
Tardis~\cite{Tardis} also uses branch-and-merge approach to weak consistency,
but does not focus on verifying the correctness of the RDTs.

Not all application logic can be expressed only using eventually consistent and
convergent RDTs. For example, a replicated bank account which guarantees
non-negative balance requires coordination between concurrent withdraw
operations. Several previous works have explored RDTs that utilize on-demand
coordination based on application invariants~\cite{KC, Gotsman, Kaki1, Cise,
Hamsaz, Ecro}. We leave the challenge of extending \name to support on-demand
coordination to future work.

\begin{acks}
We thank our shepherd, Constantin Enea, and the anonymous reviewers for their
	reviewing effort and high-quality reviews. We also thank Aseem Rastogi and
	the F* Zulip community for helping us with F* related queries.
\end{acks}


\bibliography{pldi}
\balance

\clearpage
\section*{Appendix}
\appendix
\section{Proof of Theorem 4.2}
\label{sec:proof}

\textbf{Theorem 4.2.} Given a MRDT implementation $\mathcal{D}_\tau$ of data type $\tau$, if there
	exists a valid replication-aware simulation $\mathcal{R}_{sim}$, then the
	data type implementation $\mathcal{D}_{\tau}$ is correct.

\begin{proof}

We first show that for all executions $\chi$ of the store LTS $\mathcal{M}_{D_\tau}$, $\mathcal{R}_{sim}$ holds at every transition. At every step, we will also show that the return value of every operation obeys the data-type specification and convergence modulo observable behavior is satisfied. The proof is by induction on the length of the execution.\\\\
\textbf{BASE CASE:}\\
Consider that the labelled transition system is in the initial state $C_{\bot} = (\phi_{\bot}, \delta_{\bot}, 0)$ with only one branch $b_{\bot}$.\\
$\chi = (\phi_{\bot}, \delta_{\bot}, 0)$\\
\textbf{To prove:}\\
$\mathcal{R}_{sim} (\delta (b_{\tau}), \phi (b_{\tau})) \implies (\chi \models \mathcal{F}_{\tau}$ and $\chi$ is convergent modulo observable behavior).\\\\
\textbf{Proof:}\\
For every operation o of $\tau$, let\\
$(\phi_{\bot}, \delta_{\bot}, 0) \xrightarrow{DO (o, b_{\bot})} (\phi^{'}, \delta^{'}, 1)$\\
Then according to $\Phi_{spec} (\mathcal{R}_{sim})$ in Table 2,
\begin{equation}
\label{eq:eq1}
\begin{aligned}
\mathcal{R}_{sim} (\delta_{\bot} (b_{\bot}), \phi_{\bot} (b_{\bot})) \wedge do^{\#} (\delta_{\bot} (b_{\bot}), e, o, a, 0) = I^{'} \wedge \\
 \mathcal{D}_{\tau}.do (o, \phi_{\bot} (b_{\bot}), 0) = (\phi^{'} (b_{\bot}), a)  \wedge \Psi_{ts} (\delta_{\bot} (b_{\bot}))
		\implies\\ a = \mathcal{F}_{\tau} (o, \delta_{\bot} (b_{\bot}))
\end{aligned}
\end{equation}
which is the necessary condition for $\chi \models \mathcal{F}_{\tau}$.\\

The condition for eq (2) in Section. 3 is also satisfied, since there is only one branch $b_{\bot}$ in the execution. Hence $\chi$ is convergent modulo observable behavior.
Thus the base case is proved.\\\\
\textbf{INDUCTIVE CASE:}\\
Consider an execution\\
$\chi = (\phi_{\bot}, \delta_{\bot}, 0) \xrightarrow{e_1} (\phi_{1}, \delta_{1}, t_{1}) \xrightarrow{e_2} \dots \xrightarrow{e_n} (\phi_{n}, \delta_{n}, t_{n})$.\\
\textbf{To prove:}\\
For an execution $\chi$, if $\chi \models \mathcal{F}_{\tau}$ and $\chi$ is convergent modulo observable behavior, then on applying a single step of the store execution,
the new execution obtained $\chi^{'}$, satisfies the specification and is convergent modulo observable behavior. \\\\
\textbf{Proof ($\chi^{'} \models \mathcal{F}_{\tau}$):}\\
We prove it by case-analysis on labels in the labelled transition system.\\
\textbf{Case 1:}\\
The first case is the label being CREATEBRANCH.\\
$(\phi_{n}, \delta_{n}, t) \xrightarrow{CREATEBRANCH (b_{1}, b_{2})} (\phi_{n+1}, \delta_{n+1}, t)$\\
$\phi_{n+1} = \phi_{n} [b_{2} \mapsto \phi_{n} (b_{1})] \hspace{1cm} \delta_{n+1} = \delta_{n} [b_{2} \mapsto \delta_{n} (b_{1})]$\\
$\chi^{'} = \chi \xrightarrow{CREATEBRANCH (b_{1}, b_{2})} (\phi_{n+1}, \delta_{n+1}, t)$\\
We need to prove that $\chi^{'} \models \mathcal{F}_{\tau}$.\\
For every operation o, of the data type, let\\
$(\phi_{n+1}, \delta_{n+1}, t) \xrightarrow{DO (o, b)} (\phi_{n+2}, \delta_{n+2}, t + 1)$\\
Then according to $\Phi_{spec} (\mathcal{R}_{sim})$ in Table 2,
\begin{equation}
\label{eq:eq2}
\begin{aligned}
\mathcal{R}_{sim} (\delta_{n+1} (b), \phi_{n+1} (b)) \wedge do^{\#} (\delta_{n+1} (b), e, o, a, t) = I^{'} \wedge \\
 \mathcal{D}_{\tau}.do (o, \phi_{n+1} (b), t) = (\phi_{n+2} (b), a) \wedge \Psi_{ts} (\delta_{n+1} (b))
		\implies \\a = \mathcal{F}_{\tau}  (o, \phi_{n+1} (b))
\end{aligned}
\end{equation}
which is the necessary condition for $\chi^{'} \models \mathcal{F}_{\tau}$.\\\\
\textbf{Case 2:}\\
The second case is the label being DO.\\
$(\phi_{n}, \delta_{n}, t) \xrightarrow{DO (o,b)} (\phi_{n+1}, \delta_{n+1}, t+1)$\\
$\chi^{'} = \chi \xrightarrow{DO (o,b)} (\phi_{n+1}, \delta_{n+1}, t+1)$\\
We need to prove that $\chi^{'} \models \mathcal{F}_{\tau}$.\\
For every operation o, of the data type, let\\
$(\phi_{n+1}, \delta_{n+1}, t+1) \xrightarrow{DO (o, b)} (\phi_{n+2}, \delta_{n+2}, t + 2)$\\
Then according to $\Phi_{spec} (\mathcal{R}_{sim})$ in Table 2,
\begin{equation}
\label{eq:eq3}
\begin{aligned}
\mathcal{R}_{sim} (\delta_{n+1} (b), \phi_{n+1} (b)) \wedge do^{\#} (\delta_{n+1} (b), e, o, a, t+1) = I^{'} \wedge \\
 \mathcal{D}_{\tau}.do (o, \phi_{n+1} (b), t+1) = (\phi_{n+2} (b), a) \wedge \Psi_{ts} (\delta_{n+1} (b))
		\implies \\a = \mathcal{F}_{\tau} (o, \phi_{n+1} (b))
\end{aligned}
\end{equation}
which is the necessary condition for $\chi^{'} \models \mathcal{F}_{\tau}$.\\\\
\textbf{Case 3:}\\
The second case is the label being MERGE.\\
$(\phi_{n}, \delta_{n}, t) \xrightarrow{MERGE (b_{1}, b_{2})} (\phi_{n+1}, \delta_{n+1}, t)$\\
$\chi^{'} = \chi \xrightarrow{MERGE (b_{1}, b_{2})} (\phi_{n+1}, \delta_{n+1}, t)$\\
We need to prove that $\chi^{'} \models \mathcal{F}_{\tau}$.\\
For every operation o, of the data type, let\\
$(\phi_{n+1}, \delta_{n+1}, t) \xrightarrow{MERGE (b_{1}, b_{2})} (\phi_{n+2}, \delta_{n+2}, t+1)$\\
Then according to $\Phi_{spec} (\mathcal{R}_{sim})$ in Table 2,
\begin{equation}
\label{eq:eq4}
\begin{aligned}
\mathcal{R}_{sim} (\delta_{n+1} (b), \phi_{n+1} (b)) \wedge do^{\#} (\delta_{n+1} (b), e, o, a, t) = I^{'} \wedge \\
 \mathcal{D}_{\tau}.do (o, \phi_{n+1} (b), t) = (\phi_{n+2} (b), a) \wedge \Psi_{ts} (\delta_{n+1} (b))
		\implies \\a = \mathcal{F}_{\tau} (o, \phi_{n+1} (b))
\end{aligned}
\end{equation}
which is the necessary condition for $\chi^{'} \models \mathcal{F}_{\tau}$.\\\\
\textbf{Proof ($\chi^{'}$ is convergent modulo observable behavior):}\\
Let $\chi^{'}$ be the execution obtained after applying any of the transitions (CREATEBRANCH, DO, MERGE) to $\chi$.
For proving $\chi^{'}$ is convergent modulo observable behavior, we need to show,
\begin{equation}
\label{eq:eq5}
\begin{aligned}
	\forall b_{1}, b_{2} \in dom (\phi_{i}). \delta_{i} (b_{1}) = \delta_{i} (b_{2}) \implies \phi_{i} (b_{1}) \thicksim \phi_{i} (b_{2})
\end{aligned}
\end{equation}
We know that there is a valid simulation relation between $\mathcal{D}_{\tau}$ and $\mathcal{F}_{\tau}$.
\begin{equation}
\label{eq:eq6}
\begin{aligned}
	\forall b_{1}, b_{2} \in dom (\phi_{i}).
	 \mathcal{R}_{sim} (\delta_{i} (b_{1}), \phi_{i} (b_{1})) \wedge \\ \mathcal{R}_{sim} (\delta_{i} (b_{2}), \phi_{i} (b_{2}))
\end{aligned}
\end{equation}

On substituting $\delta_{i} (b_{1})$ for $\delta_{i} (b_{2})$ according to (~\ref{eq:eq5}) in (~\ref{eq:eq6}) we get, \\
$\forall b_{1}, b_{2} \in dom (\phi_{i}). \mathcal{R}_{sim} (\delta_{i} (b_{1}), \phi_{i} (b_{1})) \wedge \\
 \mathcal{R}_{sim} (\delta_{i} (b_{2}), \phi_{i} (b_{2})) \\
\implies \mathcal{R}_{sim} (\delta_{i} (b_{1}), \phi_{i} (b_{1})) \wedge \mathcal{R}_{sim} (\delta_{i} (b_{1}), \phi_{i} (b_{2}))$\\\\
According to $\Phi_{con} (\mathcal{R}_{sim})$ in Table 2,\\
$\mathcal{R}_{sim} (\delta_{i} (b_{1}), \phi_{i} (b_{1})) \wedge \mathcal{R}_{sim} (\delta_{i} (b_{1}), \phi_{i} (b_{2})) \\
\implies \phi_{i} (b_{1}) \thicksim \phi_{i} (b_{2})$\\
Hence $\chi^{'}$ is convergent modulo observable behavior.

\end{proof}

\section{Functional queue simulation relation and implementation}
\label{sec:queue1}

\subsection{Simulation relation}
Consider the $match$ predicate as defined in Section 5.1:

\begin{equation}
	\begin{split}
		{match}_I(e_1,e_2) &  \Leftrightarrow I.oper(e_1) = enqueue(a) \\ & \wedge I.oper(e_2) = dequeue  \wedge a = I.rval(e_2)
	\end{split}
\end{equation}

We define the simulation relation, $\mathcal{R}_{sim}$ for an abstract state $I$ and a concrete state $\sigma$ as follows:
\begin{equation}
	\begin{split}
		\mathcal{R}_{sim} (I, \sigma)  \iff ((\forall (a,t) \in \sigma \iff \\
		(\exists e \in I.E \wedge I.oper(e) = enqueue(a) \wedge I.time(e) = t \wedge\\
		\neg (\exists d \in I.E \wedge {match}_I(e, d) \wedge e \xrightarrow{vis} d))) \wedge \\
		(\forall (a_1, t_1) (a_2, t_2) \in \sigma \wedge {order}_\sigma (a_1, t_1) (a_2, t_2) \iff \\ ((\exists e_1 e_2 \in I.E \wedge I.oper(e_1) = enqueue(a_1) \wedge \\ I.oper(e_2) = enqueue(a_2) \wedge I.time(e_1) = t_1 \wedge I.time(e_2) = t_2\\
		\neg (\exists d \in I.E \wedge (({match}_I(e_1, d) \wedge e_1 \xrightarrow{vis} d ) \vee \\ ({match}_I(e_2, d) \wedge e_2 \xrightarrow{vis} d))) \wedge \\ (e_1 \xrightarrow{vis} e_2 \vee (\neg(e_1 {\xrightarrow{vis}} e_2 \vee e_2 {\xrightarrow{vis}} e_1) \wedge t_1 < t_2))))))
	\end{split}
\end{equation}
where ${order}_\sigma $ $ x_1 $ $ x_2$ is the predicate that states $x_1$ occurs before $x_2$ in the concrete state $\sigma$.

The simulation relation $\mathcal{R}_{sim}$ consists of two parts. The first part states that for any element $a$ in the concrete state of the queue, there exists an enqueue operation, $e$ in the abstract state, that is not matched with any dequeue operation, and the converse of this. The second part of the relation states that for any two elements $a_1$ and $a_2$ in the concrete state of the queue, such that $a_1$ occurs before $a_2$, there exist two enqueue operations, $e_1$ and $e_2$ in the abstract state, that are not matched with any dequeue operation, such that $e_1 \xrightarrow{vis} e_2$ ($e_2$ was performed after $e_1$)  or $ (\neg(e_1 {\xrightarrow{vis}} e_2 $ $ \vee $ $e_2 {\xrightarrow{vis}} e_1) $ $\wedge$ $ t_1 < t_2)$ ($e_1$ and $e_2$ are concurrent operations), and the converse of this. The first part takes care of the membership of the elements in the queue, while the second part takes care of the ordering of the elements in the queue.

\subsection{Functional queue implementation}
The definition of the state of the functional queue is as follows:
\begin{lstlisting}
	type s =
		| S of (int * int) list *
					(int * int) list
\end{lstlisting}

The enqueue and dequeue functions for the functional queue are defined as follows:
\begin{lstlisting}
	let enqueue x q =
		(S q.front (x::q.back))
\end{lstlisting}

\begin{lstlisting}
	let norm q =
	match q with
	|(S [] back) -> (S (rev back) [])
	|_ -> q
\end{lstlisting}

\begin{lstlisting}
	let dequeue q =
	match q with
	|(S [] []) -> (None, q)
	|(S (x::xs) _) -> (Some x, (S xs q.back))
	|(S [] (x::xs)) ->
	  let (S (y::ys) []) = norm q in
	(Some y, (S ys []))
\end{lstlisting}

Following are the definitions of some functions used as helper functions for the three-way merge. In all of these definitions, the first element of the pair is taken to be the timestamp, and the second element to be the actual enqueued element. \emph{union1} is used to merge two lists of pairs (with unique first elements) that are sorted according to the first element of the pair:
\begin{lstlisting}
	let rec union l1 l2 =
	match l1, l2 with
	| [], [] -> []
	| [], l2 -> l2
	| l1, [] -> l1
	| h1::t1, h2::t2 -> if (fst h1 < fst h2)
	then h1::(union t1 l2)
		else h2::(union l1 t2)
\end{lstlisting}

\emph{diff\_s} is used to find the difference between two lists of pairs with unique first elements, that are sorted according to the first element of the pair. Additionally, in the context where \emph{diff\_s} is used, \emph{a} is a child of \emph{l}. This function is used to find the newly enqueued elements in \emph{a}. This simplifies the function to finding the longest contiguous subsequence that is present in \emph{a} but not \emph{l}. This also can be interpreted as finding the longest suffix of \emph{a} that is not present in \emph{l}, since all the newly enqueued elements occur after the existing elements. This task can be completed in $O(n)$ time where $n$ is the length of the longest of the two lists.
\begin{lstlisting}
	let rec diff_s a l =
	match a, l with
	| x::xs, y::ys -> if (fst y) < (fst x)
		then diff_s (x::xs) ys else (diff_s xs ys)
	| [], y::ys -> []
	| _, [] -> a
\end{lstlisting}

\emph{intersection} is used to find the longest common contiguous subsequence between \emph{l}, \emph{a} and \emph{b}. Again, in the context where \emph{intersection} is used, \emph{a} and \emph{b} are children of \emph{l}. This finds the list of elements that have not been dequeued in either \emph{a} or \emph{b}. Hence, the problem is simplified to finding the longest contiguous subsequence of \emph{l} that is a prefix of \emph{a} and \emph{b}. Since all the three lists are sorted according to the first element of the pair, this task can be completed in $O(n)$ time where $n$ is the length of the longest of the three lists.
\begin{lstlisting}
	let rec intersection l a b =
	match l, a, b with
	| x::xs, y::ys, z::zs ->
		if ((fst x) < (fst y)
			|| (fst x) < (fst z)) then
		(intersection xs (y::ys) (z::zs))
		else (x::(intersection xs ys zs))
	| x::xs, [], z::zs -> []
	| x::xs, y::ys, [] -> []
	| x::xs, [], [] -> []
	| [], _, _ -> []
\end{lstlisting}

\emph{tolist} is used to convert a functional queue to a single list. This function takes $O(n)$ time where $n$ is the length of the functional queue.
\begin{lstlisting}
	let tolist (S f b) = append f (rev b)
\end{lstlisting}

For the three-way merge between \emph{l}, \emph{a} and \emph{b}, we first find the elements that are not dequeued in \emph{a} or \emph{b}. Then we find the list of newly enqueued elements in \emph{a} and \emph{b}, and append it to the list of undequeued elements. The core of the three-way merge is defined as follows:
\begin{lstlisting}
	let merge_s l a b =
		let ixn = intersection l a b in
		let diff_a = diff_s a l in
		let diff_b = diff_s b l in
		let union_ab = union diff_a diff_b in
			append ixn union_ab
\end{lstlisting}
The three-way merge for \emph{l}, \emph{a} and \emph{b} is defined as follows:

\begin{lstlisting}
	let merge l a b =
		(S
		(merge_s (tolist l) (tolist a) (tolist b))
			[])
\end{lstlisting}
Since all the tasks involved in \emph{merge} take linear time in terms of the length of the longest list, which is equal to that of the longest queue, \emph{merge} takes $O(n)$ time where $n$ is the length of the longest queue.

\end{document}